\documentclass[sigconf]{acmart}

\usepackage{multirow}
\usepackage{enumitem}

\definecolor{gruvred}{HTML}{CC241D}
\definecolor{gruvyellow}{HTML}{D79921}
\definecolor{gruvblue}{HTML}{458588}
\definecolor{gruvgreen}{HTML}{98971A}

\DeclareRobustCommand{\M}{\textup{MELODI}}
\AtEndPreamble{%
  \theoremstyle{acmdefinition}%
  \newtheorem{remark}[theorem]{Remark}}

\renewcommand{\dbltopfraction}{0.9}

\copyrightyear{2027}
\acmYear{2027}
\setcopyright{cc}
\setcctype{by}
\acmConference[WSDM '27]{Proceedings of the Twentieth ACM International Conference on Web Search and Data Mining}{February 15--19, 2027}{Hong Kong, Hong Kong}
\acmBooktitle{Proceedings of the Twentieth ACM International Conference on Web Search and Data Mining (WSDM '27), February 15--19, 2027, Hong Kong, Hong Kong}
\acmDOI{10.1145/3862322.3863572}
\acmISBN{979-8-4007-3117-4/2027/02}

\begin{document}
\settopmatter{authorsperrow=4}

\title[MELODI]{\texorpdfstring{MELODI: Similarity--Identity Decoupling\\for Lightweight Ranking}{MELODI: Similarity--Identity Decoupling for Lightweight Ranking}}

\author{Quoc Phong Nguyen}
\orcid{0000-0002-0183-2525}
\affiliation[obeypunctuation=true]{%
  \department{Amazon International Machine Learning}, \institution{Amazon}\\
  \country{Australia}}
\email{qphong@amazon.com}

\author{Vuong Le}
\orcid{0000-0003-1582-1269}
\affiliation[obeypunctuation=true]{%
  \department{Amazon International Machine Learning}, \institution{Amazon}\\
  \country{Australia}}
\email{levuong@amazon.com}

\author{Volodymyr Vaskovych}
\orcid{0000-0002-1075-4549}
\affiliation[obeypunctuation=true]{%
  \department{Amazon International Machine Learning}, \institution{Amazon}\\
  \country{Australia}}
\email{vaskovyc@amazon.com}

\author{Julien Monteil}
\orcid{0000-0002-8723-2677}
\affiliation[obeypunctuation=true]{%
  \department{Amazon International Machine Learning}, \institution{Amazon}\\
  \country{Australia}}
\email{jul@amazon.com}

\renewcommand{\shortauthors}{Nguyen et al.}

\begin{abstract}
For latency-constrained candidate ranking, we propose a \emph{similarity--identity decoupling} framework that separates target--history compatibility from target-only information. We study two orders for computing pooled similarity, \emph{similarity-then-pool} and \emph{pool-then-similarity}, and identify linearity of the similarity operator in its history argument, for each fixed target, as sufficient for their equivalence, justifying the cheaper pool-then-similarity form. We instantiate the framework as \emph{multi-head efficient lightweight operator for decoupling identity and similarity}~(\M{}), whose similarity branch combines target-conditioned multi-head history pooling with a single residual-free cross layer that is linear in its pooled-history input and whose identity branch is problem-dependent. On public benchmark datasets, \M{} matches or surpasses the evaluated target-aware and target-agnostic baselines. On two large-scale industrial datasets, its \emph{area under the precision--recall curve}~(AUPRC), \emph{normalised discounted cumulative gain}~(NDCG), and recall values are equal to or higher than those of the evaluated target-aware baselines. Processor measurements show lower latency than the evaluated candidate-conditioned baselines at large candidate counts, further reduced by request-local history reuse.
\end{abstract}




\maketitle

\section{Introduction}

Large-scale recommenders commonly separate candidate generation from ranking: candidate generation reduces millions of items to hundreds or thousands, after which a richer model ranks the shortlist~\citep{covington2016deep,yi2019sampling}. Candidate generation can use separately encoded user or query and item representations with a simple similarity scorer~\citep{huang2013dssm,yi2019sampling}. This decomposition permits item representations to be precomputed and searched efficiently at catalogue scale~\citep{yi2019sampling}. By contrast, ranking operates on the smaller candidate set and can afford candidate-specific \emph{early fusion}, which jointly processes each shortlisted candidate with the user's history and request features~\citep{zhou2018deep,chen2019bst}, although real-time serving still constrains the computation performed for each candidate.

Target-aware encoders use this per-candidate budget to distinguish a user's concurrent interests. Consider a shopper whose recent activity alternates between running gear and photography equipment. When scoring running shoes, the model should emphasise the sports interactions, and when scoring a camera lens, the photography interactions. A candidate-independent vector represents both interests jointly and may not identify which events support each candidate. Target-\emph{agnostic} models such as the \emph{Self-Attentive Sequential Recommender}~(SASRec)~\citep{kang2018sasrec} reuse one history summary for every candidate. Target-aware models such as the \emph{Deep Interest Network}~(DIN)~\citep{zhou2018deep}, \emph{Deep Interest Evolution Network}~(DIEN)~\citep{zhou2019dien}, \emph{Deep Session Interest Network}~(DSIN)~\citep{feng2019dsin}, and \emph{Search-based Interest Model}~(SIM)~\citep{pi2020sim} instead condition part of their computation on each candidate. Target-aware Transformers such as the \emph{Behavior Sequence Transformer}~(BST) jointly encode each candidate with the history, requiring one encoder evaluation per candidate~\citep{chen2019bst}. DIN evaluates a \emph{multilayer perceptron}~(MLP) for every (history item, candidate) pair. Thus, a first requirement is to keep per-candidate computation lightweight as histories and candidate sets grow.

Candidate conditioning determines which history events matter for each target, but it does not remove a second modelling requirement: target-aware ranking must preserve target identity and user--target compatibility as distinct signals. A target's \emph{identity} carries target-only information such as base-rate appeal and within-target feature interactions, whereas its \emph{similarity} to the relevant history represents user--target compatibility. For example, a holiday decoration may become broadly appealing during Christmas because the scorer can combine its target-only identity with calendar context, without requiring a matching seasonal pattern in the shopper's history. Compatibility is a different relation: a drill may score highly for a shopper whose recent activity includes paint, fixtures, and tools because it matches that renovation history. DIN can supply the target separately to its scorer, but its weighted pooling combines target-conditioned history content and compatibility in one vector. We instead decouple these roles: a similarity branch computes an explicit user--target interaction feature, an identity branch processes the target alone, and the scorer combines both signals with any contextual features.

We propose a target-aware ranking framework that separates user--target similarity from target identity and retains a reusable history path through three contributions:
\begin{itemize}[wide=0pt,nosep]
\item \textbf{Similarity--identity framework.} We formalise similarity-then-pool and pool-then-similarity (SP/PS) and state the elementary condition that justifies our design: when the similarity operator is linear in its history argument, the cheaper PS order exactly recovers the SP output after target-aware pooling (Prop.~\ref{prop:sp-ps-equivalence}).
\item \textbf{\M{} architecture.} We instantiate the framework as \emph{multi-head efficient lightweight operator for decoupling identity and similarity}~(\M{}). Its similarity branch combines multi-head dot-product cross-attention~\citep{vaswani2017attention} with a single cross layer based on the \emph{Deep \& Cross Network v2}~(DCN-v2)~\citep{wang2021dcn}. Both primitives exist; the new elements are a two-input cross that is linear in the pooled history and has no residual, per-head gating over shared experts, and a separate identity branch, which together keep the history path target-independent and reusable across candidates (Sec.~\ref{sec:architecture}).
\item \textbf{Evaluation across public and industrial settings.} We evaluate ranking quality, component variants, analytical \emph{floating-point operation}~(FLOP) counts, and processor latency on two public and two industrial datasets (Sec.~\ref{sec:experiments}, Supp.~Secs.~\ref{app:flops} and~\ref{sec:ablations}). The prespecified public configuration records higher values than every baseline in Tab.~\ref{tab:main} on every reported metric. In both settings, \M{} offers lower processor latency than the evaluated candidate-conditioned baselines for large candidate sets, while DIN remains faster up to $C=500$ in the public benchmark. The component comparisons delimit which defaults the measurements support.
\end{itemize}

\section{Related Work}

\noindent\textbf{Sequential models for ranking.} Within the downstream ranking stage, reusable target-independent encoders such as GRU4Rec~\citep{hidasi2016gru4rec}, SASRec~\citep{kang2018sasrec}, and BERT4Rec~\citep{sun2019bert4rec} produce one history representation and confine candidate interaction to the scorer. The \emph{Multi-channel User Interest Memory Network} likewise maintains candidate-independent long-history memory and avoids request-time interaction between the candidate and behaviour sequence~\citep{pi2019mimn}.

Candidate-conditioned pooling instead changes the history representation for each target. DIN evaluates an MLP activation for every (history item, candidate) pair and uses scalar weights to pool one target-aware history~\citep{zhou2018deep}. DIEN evolves interests with a target-conditioned gated recurrent unit~\citep{zhou2019dien}, DSIN activates session representations after self-attention and a bidirectional long short-term memory network~\citep{feng2019dsin}, and SIM performs target-conditioned general and exact search~\citep{pi2020sim}. The \emph{TWo-stage Interest Network}~(TWIN) targets lifelong sequences of $10^4$--$10^5$ behaviours, far beyond our 50, and caches projections of item-inherent features to make target attention affordable~\citep{chang2023twin}. AutoAttention replaces DIN's MLP activation with dot products over raw field embeddings and learns globally shared field-pair weights that are pruned during training~\citep{zheng2022autoattention}. The \emph{Co-Action Network}~(CAN) applies a target-parameterised MLP to each behaviour and sum-pools the outputs~\citep{bian2022can}, a similarity-then-pool operator that is non-linear in the history item.

Joint target--history encoders move candidate conditioning inside the sequential model. BST applies self-attention to a combined target--history sequence, so the full encoder is re-run per candidate~\citep{chen2019bst}. For ranking, \emph{Hierarchical Sequential Transduction Units}~(HSTU) score candidate microbatches in one pass, masking attention between candidates and caching history keys and values~\citep{zhai2024hstu}. Our proposed \M{} also reuses history keys and values, but through one cross-attention layer over a history encoded without self-attention, with target-dependent per-head gating and a cross that is linear in its pooled-history input.

\noindent\textbf{Feature interaction and similarity--identity decoupling.} \emph{Factorization Machines}~\citep{rendle2010fm}, Wide\&Deep~\citep{cheng2016wide}, DeepFM~\citep{guo2017deepfm}, and DCN-v2~\citep{wang2021dcn} operate on flat features. We adapt DCN-v2-style cross layers with two inputs $(\mathbf{u}, \mathbf{t})$, block-diagonal experts, and per-head gating. Related methods modify different stages: the \emph{Feature Generation by Convolutional Neural Network}~(FGCNN)~\citep{liu2019fgcnn} generates interactions by convolving over flat feature embeddings, while the decoupled-embedding approach of \citet{feng2025dare} separates attention and representation embedding tables but still aggregates the retrieved behaviours into one representation. In our proposed \M{}, the scorer instead receives an explicit user--target interaction vector and a separate target-only vector.

\section{Similarity--Identity Decoupling: Framework}
\label{sec:framework}

\subsection{Problem Setup}

We study the ranking stage and take its candidate set as given rather than modelling candidate generation. The ranker scores each candidate against the same recent interaction history under a per-request latency budget. Let the history be $\mathbf{X} = [\mathbf{x}_1, \ldots, \mathbf{x}_S]^\top \in \mathbb{R}^{S \times D_x}$ and the candidate set be $\mathcal{C} = \{\mathbf{t}_1, \ldots, \mathbf{t}_C\} \subset \mathbb{R}^{D_t}$. Keeping $D_x$ and $D_t$ distinct allows the framework to cover similarity operators whose pooled-history and target inputs have different widths. The ranker produces one scalar per candidate.

\begin{table*}[t]
\centering
\scriptsize
\caption{Mathematical notation used in the framework and architecture.}
\label{tab:notation}
\setlength{\tabcolsep}{2pt}
\renewcommand{\arraystretch}{1.08}
\begin{tabular}{@{}p{3.25cm}p{5.2cm}p{3.25cm}p{5.2cm}@{}}
\toprule
Symbol & Description & Symbol & Description \\
\midrule
\multicolumn{4}{@{}l}{\textit{Framework}} \\
$\mathcal{C}=\{\mathbf{t}_c\}_{c=1}^{C}$ & Candidate set of size $C$ &
$\mathbf{X}=[\mathbf{x}_i]_{i=1}^{S}\in\mathbb{R}^{S\times D_x}$ & History matrix of length $S$ \\
$\mathbf{x}_i\in\mathbb{R}^{D_x},\ \mathbf{t}\in\mathbb{R}^{D_t}$ & History item and generic target; the candidate index on $\mathbf{t}$ is dropped &
$D_x,D_t$ & Framework history and target widths \\
$\boldsymbol{\alpha}=[\alpha_i]_{i=1}^{S}\in\Delta^{S-1}$ & Non-negative attention distribution over the $S$ history items &
$\mathbf{u}(\mathbf{X},\mathbf{t})\in\mathbb{R}^{D_x}$ & Target-aware pooled history \\
$\phi:\mathbb{R}^{D_x}\!\times\!\mathbb{R}^{D_t}\!\to\!\mathbb{R}^{M}$ & Similarity operator; linear in its history argument where SP/PS equivalence is invoked (Def.~\ref{def:history-linear}) &
$M$ & Similarity-operator output width \\
$\boldsymbol{\psi}_{\mathrm{SP}},\boldsymbol{\psi}_{\mathrm{PS}}\in\mathbb{R}^{M}$ & Pooled similarity under similarity-then-pool and pool-then-similarity orderings &
$\mathbf{A}(\mathbf{t})\in\mathbb{R}^{M\times D_x}$ & Matrix in $\phi_{\mathbf A}(\mathbf{x},\mathbf{t})=\mathbf{A}(\mathbf{t})\mathbf{x}$ \\
$\boldsymbol{\eta}:\mathbb{R}^{D_t}\!\to\!\mathbb{R}^{D_x}$ & Target map in $\phi(\mathbf{x},\mathbf{t})=\mathbf{x}\odot\boldsymbol{\eta}(\mathbf{t})$ &
$\odot$ & Element-wise product \\
$\mathbf{M},\mathbf{M}(\mathbf{t})\in\mathbb{R}^{D_x\times D_t}$ & Linear and target-dependent maps in the framework examples &
$\mathbf{U}_k\in\mathbb{R}^{D_x\times r},\mathbf{V}_k\in\mathbb{R}^{r\times D_t}$ & Framework factors for expert $k$; $\mathbf{M}(\mathbf{t})$ mixes them using $g_k(\mathbf{t})$ \\
\midrule
\multicolumn{4}{@{}l}{\textit{\M{} architecture}} \\
$B$ & Batch size &
$D_{\mathrm{raw}}$ & Raw pretrained embedding width \\
$D$ & Item-embedding width after Stage 1 &
$H,d$ & Number of attention heads and per-head width \\
$D_u=Hd$ & Concatenated attention and similarity width; need not equal $D$ &
$K,r$ & Number of low-rank experts and expert rank \\
$\mathbf{X}^{\mathrm{raw}},\mathbf{T}^{\mathrm{raw}}$ & Raw history and candidates in $\mathbb{R}^{S\times D_{\mathrm{raw}}}$ and $\mathbb{R}^{C\times D_{\mathrm{raw}}}$ &
$\tilde{\mathbf{X}},\tilde{\mathbf{x}}_i,\tilde{\mathbf{t}}$ & Projected, normalised history, history item, and target at width $D$ \\
$\mathbf{W}_q,\mathbf{W}_k,\mathbf{W}_v,\mathbf{W}_{g,h}$ & Query, key, shared value, and per-head gate projection matrices &
$\mathbf{Q},\mathbf{K},\mathbf{V}$ & Projected query, key, and value tensors; subscript $h$ selects a head \\
$\boldsymbol{\alpha}_h\in\mathbb{R}^{S}$ & Attention weights for one candidate in head $h$; batched shape is $C\times S$ &
$\mathbf{u}_h\in\mathbb{R}^{d}$ & Per-candidate pooled history for head $h$ \\
$\mathbf{u}=[\mathbf{u}_1;\ldots;\mathbf{u}_H]\in\mathbb{R}^{D_u}$ & Concatenated pooled history &
$\mathbf{v}_t,\mathbf{v}_t^{(h)}$ & Target under $\mathbf{W}_v$ in $\mathbb{R}^{D_u}$, and its head-$h$ slice in $\mathbb{R}^{d}$ \\
$\mathbf{U}_k\in\mathbb{R}^{d\times r},\mathbf{V}_k\in\mathbb{R}^{r\times d}$ & \M{}'s factors for expert $k$; their product has rank at most $r$ &
$\mathbf{g}_h\in\mathbb{R}^{K},g_{h,k}$ & Gate vector for head $h$ and its expert-$k$ component \\
$\boldsymbol{\eta}_h,\mathbf{b}_h\in\mathbb{R}^{d}$ & Per-head target-map output and bias; Stage 3 uses $\boldsymbol{\eta}:\mathbb{R}^{D}\to\mathbb{R}^{D_u}$ &
$\mathbf{c}_h\in\mathbb{R}^{d}$ & Per-head similarity output \\
$\mathbf{c}\in\mathbb{R}^{D_u}$ & Concatenated similarity signal &
$\tilde{\mathbf{t}}_{\mathrm{id}}\in\mathbb{R}^{D}$ & Identity-branch target representation \\
\midrule
\multicolumn{4}{@{}l}{\textit{Serving and baselines}} \\
$T_{\mathrm{shared}},T_{\mathrm{per\mbox{-}cand}}$ & History-only and per-candidate serving costs &
$\mathbf{h}$ & History-item representation in the DIN activation unit \\
\bottomrule
\end{tabular}
\end{table*}

The design problem is to construct an intermediate representation that is informative but cheap to evaluate for every candidate. We separate it into a \emph{similarity} signal, which depends on both $\mathbf{X}$ and $\mathbf{t}_c$, and an \emph{identity} signal, which depends only on $\mathbf{t}_c$. The similarity branch is the focus of Secs.~\ref{sec:similarity-branch}--\ref{sec:examples}. The identity branch\label{sec:identity-branch} may use the raw target embedding, a target MLP, a DCN-v2 target cross, or their combination to capture target-intrinsic information such as base-rate appeal and within-target feature interactions.

The framework is agnostic to the embedding source. Item representations may come from content encoders, collaborative filtering, or self-supervised sequential models. A non-linear adapter may be applied independently to each raw history or target embedding before the similarity operator. Definition~\ref{def:history-linear} then applies to the adapted vectors. User, context, and other request-level features may enter the final scorer or a separate feature-crossing module without changing whether the similarity branch is history-linear.

\subsection{The Similarity Branch: Two Constructions of the Pooled Similarity}
\label{sec:similarity-branch}

For a given target $\mathbf{t} \in \mathbb{R}^{D_t}$ (we drop the candidate index $c$ hereafter), let $\boldsymbol{\alpha} \in \Delta^{S-1}$ denote a non-negative attention weight vector satisfying $\sum_i \alpha_i = 1$. The weights $\boldsymbol{\alpha}$ may depend on $(\mathbf{X}, \mathbf{t})$, and the analysis allows any attention function. Let $\phi : \mathbb{R}^{D_x} \times \mathbb{R}^{D_t} \to \mathbb{R}^M$ denote a \emph{similarity operator}. We define \emph{similarity-then-pool}~(SP) and \emph{pool-then-similarity}~(PS) as
\begin{align}
    \text{(SP):} \quad
    \boldsymbol{\psi}_{\text{SP}}(\mathbf{X}, \mathbf{t}) &\;\triangleq\; \textstyle\sum_{i=1}^{S} \alpha_i \, \phi(\mathbf{x}_i, \mathbf{t}) \;\in\; \mathbb{R}^M,
    \label{eq:sim-then-pool} \\
    \text{(PS):} \quad
    \boldsymbol{\psi}_{\text{PS}}(\mathbf{X}, \mathbf{t}) &\;\triangleq\; \phi\bigl(\mathbf{u}(\mathbf{X},\mathbf{t}),\; \mathbf{t}\bigr) \;\in\; \mathbb{R}^M,
    \label{eq:pool-then-sim}
\end{align}
where $\mathbf{u}(\mathbf{X}, \mathbf{t}) \triangleq \sum_{i=1}^S \alpha_i \mathbf{x}_i \in \mathbb{R}^{D_x}$ is the \emph{target-aware pooled history vector}. SP applies $\phi$ before aggregation, whereas PS first forms $\mathbf{u}$ and invokes $\phi$ once. We write both arguments of $\mathbf{u}(\mathbf{X}, \mathbf{t})$ explicitly because target-dependent attention changes the pooled history for the candidate being scored.

\noindent\textbf{DIN within the framework.} At its history-pooling stage, DIN corresponds to PS with $\phi(\mathbf{u}, \mathbf{t}) \triangleq \mathbf{u}$. This identity operator is history-linear under Definition~\ref{def:history-linear}, but DIN is not linear in the raw history end to end. Its activation unit computes non-linear, target-dependent attention weights for every history item, and its downstream scorer may also be non-linear. The SP/PS condition constrains neither component, so it does not remove DIN's per-item activation cost. Because $\mathbf{u}$ combines history content with target-conditioned attention weights, compatibility remains implicit rather than appearing as a separate feature. DIN also assigns one scalar weight to each history item, giving all dimensions the same target-conditioned attention distribution. \M{} instead uses $H$ attention heads (Sec.~\ref{sec:cross-attn}) to learn distinct pooling distributions before applying an explicit target-dependent history-linear similarity operator $\phi$ (Sec.~\ref{sec:examples}) to each target-aware pooled history.

\begin{remark}[Fidelity--efficiency tension between the two constructions]
\label{rem:fidelity-efficiency}
The two constructions are in tension. \textbf{SP is preferred for modelling fidelity}: $\phi$ sees each history item $\mathbf{x}_i$ in isolation and compares it to the target before any aggregation, so the output reflects per-item similarity. \textbf{PS is preferred for computational efficiency}: it requires one operator evaluation per (user, target) pair rather than $S$, reducing the evaluation count by a factor of $S$. This tension motivates identifying families of operators for which \emph{both} orderings produce the same output, so that the cheaper PS form reproduces the more faithful SP signal.
\end{remark}

\noindent\textbf{Sufficient condition for SP/PS equivalence.} The required condition concerns only the first, history argument of $\phi$ and places no restriction on how $\phi$ depends on $\mathbf{t}$.

\begin{definition}[History-linear similarity]
\label{def:history-linear}
For each fixed $\mathbf{t}$, $\phi$ is \emph{history-linear} when $\phi(\cdot,\mathbf{t})$ is linear in its history argument.
\end{definition}

This property applies only to the similarity operator $\phi$. It does not require the attention weights $\boldsymbol{\alpha}$, the pooled-history map $\mathbf{u}(\mathbf{X},\mathbf{t})$, any embedding adapter, or the final scorer to be linear in the raw interaction history $\mathbf{X}$ as a whole.

\begin{proposition}[SP/PS equivalence under history-linearity]
\label{prop:sp-ps-equivalence}
For any history $\mathbf{X}$, target $\mathbf{t}$, and attention distribution $\boldsymbol{\alpha}\in\Delta^{S-1}$, a history-linear similarity operator satisfies
$\boldsymbol{\psi}_{\mathrm{SP}}(\mathbf{X},\mathbf{t})=
\boldsymbol{\psi}_{\mathrm{PS}}(\mathbf{X},\mathbf{t})$.
\end{proposition}

\begin{proof}
For fixed $\mathbf{t}$, linearity in the history argument gives
\begin{equation}
    \boldsymbol{\psi}_{\mathrm{SP}}
    =\textstyle\sum_{i=1}^{S}\alpha_i\phi(\mathbf{x}_i,\mathbf{t})
    =\phi\!\left(\textstyle\sum_{i=1}^{S}\alpha_i\mathbf{x}_i,\mathbf{t}\right)
    =\boldsymbol{\psi}_{\mathrm{PS}}.
    \label{eq:linear-first}
\end{equation}
\end{proof}

Proposition~\ref{prop:sp-ps-equivalence} permits candidate-dependent attention and an arbitrarily non-linear target map because only the history slot must remain linear. The PS order can therefore recover the SP pooled similarity with one invocation of $\phi$ per candidate. The proposition follows directly from linearity, and its role is architectural: it identifies the similarity operators that admit this exact reordering.

\subsection{Examples of History-Linear Similarity Operators}
\label{sec:examples}

The most general history-linear operator has the form $\phi_{\mathbf{A}}(\mathbf{x}, \mathbf{t}) = \mathbf{A}(\mathbf{t})\, \mathbf{x}$ for any matrix-valued function $\mathbf{A} : \mathbb{R}^{D_t} \to \mathbb{R}^{M \times D_x}$. Target information enters only through $\mathbf{A}(\mathbf{t})$, so $\sum_i \alpha_i \mathbf{A}(\mathbf{t}) \mathbf{x}_i = \mathbf{A}(\mathbf{t}) \mathbf{u}$ regardless of how $\mathbf{A}$ depends on $\mathbf{t}$. We focus on the element-wise specialisation with $M=D_x$,
\begin{equation}
    \phi(\mathbf{x}, \mathbf{t}) \;=\; \mathbf{x} \odot \boldsymbol{\eta}(\mathbf{t}) \;\in\; \mathbb{R}^{D_x},
    \label{eq:general-cross}
\end{equation}
where $\mathbf{A}(\mathbf{t}) = \mathrm{diag}(\boldsymbol{\eta}(\mathbf{t}))$ for a target map $\boldsymbol{\eta} : \mathbb{R}^{D_t} \to \mathbb{R}^{D_x}$. The four cases below increase the map's complexity.

\noindent\textbf{(1) Pass-through target map: $\boldsymbol{\eta}(\mathbf{t}) = \mathbf{t}$.} When $D_x=D_t$, $\phi(\mathbf{x}, \mathbf{t}) = \mathbf{x} \odot \mathbf{t}$, whose element-wise sum recovers the dot product $\mathbf{u}^\top \mathbf{t}$. The downstream MLP can instead receive the full vector $\mathbf{u} \odot \mathbf{t}$ and learn how to weight its dimensions.

\noindent\textbf{(2) Linear target map: $\boldsymbol{\eta}(\mathbf{t}) = \mathbf{M}\mathbf{t}$.} $\phi_{\text{cross}}(\mathbf{x}, \mathbf{t}) = \mathbf{x} \odot (\mathbf{M}\mathbf{t})$ with learnable $\mathbf{M} \in \mathbb{R}^{D_x \times D_t}$. Summing its elements gives the bilinear form $\mathbf{u}^\top \mathbf{M} \mathbf{t}$.

\noindent\textbf{(3) Input-dependent mixture (\M{}'s choice): $\boldsymbol{\eta}(\mathbf{t}) = \mathbf{M}(\mathbf{t})\,\mathbf{t}$.} With $\mathbf{M}(\mathbf{t}) = \sum_k g_k(\mathbf{t}) \mathbf{U}_k \mathbf{V}_k$, the target-dependent matrix is a softmax-gated mixture of $K$ low-rank experts $(\mathbf{U}_k, \mathbf{V}_k)$ with gating weights $g_k(\mathbf{t})$, inspired by the low-rank mixture-of-experts cross layer of DCN-v2~\citep{wang2021dcn}. Despite the softmax non-linearity, $\boldsymbol{\eta}$ acts only on $\mathbf{t}$, so the operator remains history-linear and the PS equivalence is preserved. We apply the operator independently to each of the $H$ attention heads, with per-head gating weights $g_{h,k}(\mathbf{t})$, so each head produces its own target-dependent mixture and similarity vector. The architectural realisation is detailed in Sec.~\ref{sec:per-head-cross}.

\noindent\textbf{(4) Arbitrary non-linear target map, e.g., $\boldsymbol{\eta} = \mathrm{MLP}$.} Any non-linear function of $\mathbf{t}$ alone preserves history-linearity because $\boldsymbol{\eta}(\mathbf{t})$ is constant across the sum over history items.

\begin{remark}[Expressiveness--efficiency trade-off]
\label{rem:expr-eff}
Allowing $\phi$ to depend non-linearly on its history argument enlarges the operator class, but SP and PS then no longer agree in general because the non-linearity does not commute with pooling. Preserving their equivalence therefore restricts $\phi$, rather than the complete model, to be history-linear. A non-linear operator that realises the SP semantics must instead be invoked once per history item. This restriction trades expressiveness in $\phi$'s history slot for the cheaper PS form, which suits latency-constrained ranking (Sec.~\ref{sec:production}).
\end{remark}

\begin{figure*}[t]
\centering
\includegraphics[width=0.8\textwidth]{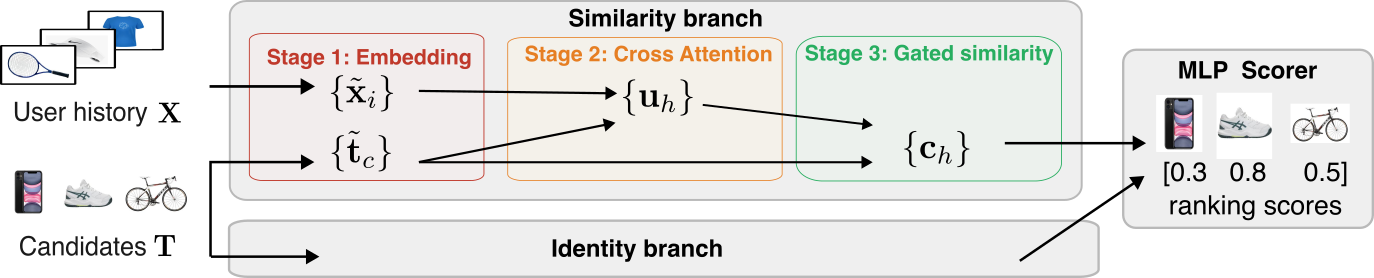}
    \caption{\M{}'s two-signal design. Top row: the similarity signal branch applies a shared embedding projection and \emph{layer normalization}~(LN)~\citep{ba2016layer} (Stage 1), multi-head dot-product cross-attention producing $\{\mathbf{u}_h\}$ (Stage 2), and a per-head gated cross layer without a residual connection producing $\mathbf{c}$ (Stage 3). Bottom row: the identity signal branch (dashed) applies a problem-dependent target-only transformation.}
    \Description{Block diagram of MELODI's two-signal design. The upper similarity
      signal branch runs left to right: the user history and the target item both enter a
      shared embedding and layer-normalization stage, then multi-head dot-product
      cross-attention pools the history into per-head vectors, then a per-head
      gated cross layer without a residual connection combines those with the
      target to produce the similarity signal c. The lower, dashed identity signal branch
      applies a target-only transformation. Both signal outputs enter a final MLP
      that emits the score.}
\label{fig:architecture}
\end{figure*}

\section{The \M{} Architecture}
\label{sec:architecture}

\begin{figure*}[t]
\centering
\includegraphics{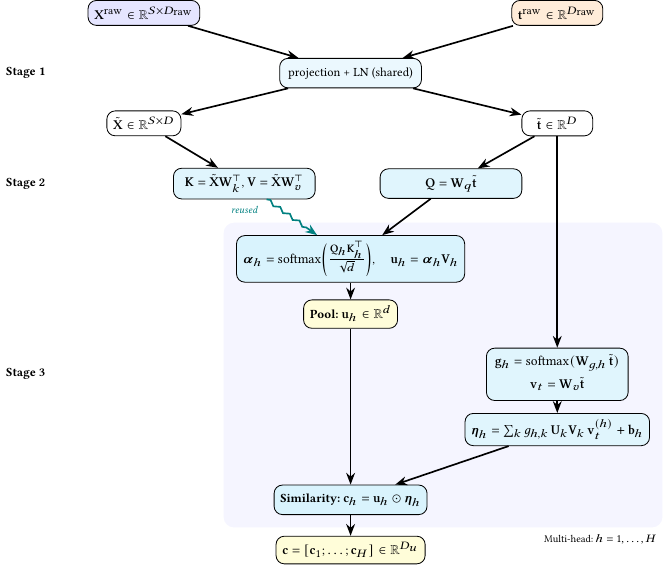}
\caption{Detailed data flow of the similarity branch. Stage 1 produces $\tilde{\mathbf{X}}$ and $\tilde{\mathbf{t}}$ at width $D$. Stage 2 projects them to $H$ heads of width $d$ and computes the pooled histories $\mathbf{u}_h$. Stage 3 projects the target to $\mathbf{v}_t\in\mathbb{R}^{D_u}$, where $D_u=Hd$, and applies per-head gated experts to produce $\boldsymbol{\eta}_h$ and $\mathbf{c}_h$. The teal wavy arrows mark the cacheable history-only keys and values (Rem.~\ref{rem:efficiency}).}
\Description{Vertical data-flow diagram of the similarity branch in three
  stages. Stage 1: the history matrix and the target vector both pass through a
  shared projection and layer normalization. Stage 2: the history produces keys and values
  while the target produces the query. Scaled dot-product cross-attention then
  yields a pooled history vector per head. The key and value boxes are marked
  reused, indicating they depend on the history only and are computed once per
  request. Stage 3: per-head softmax-gated low-rank experts map the target to a
  vector that is multiplied element-wise with the pooled history, and the
  per-head results are concatenated into the similarity signal c.}
\label{fig:similarity-detail}
\end{figure*}

Our proposed \M{} takes a raw interaction history $\mathbf{X}^{\mathrm{raw}} \in \mathbb{R}^{S \times D_{\mathrm{raw}}}$ and $C$ raw candidate embeddings $\mathbf{T}^{\mathrm{raw}} \in \mathbb{R}^{C \times D_{\mathrm{raw}}}$ and produces one score per candidate. Its similarity and identity branches instantiate the framework of Sec.~\ref{sec:framework}. Fig.~\ref{fig:architecture} shows the overall design. The similarity branch comprises Stages 1--3 below, with data flow in Fig.~\ref{fig:similarity-detail} and full equations in Supp.~Sec.~\ref{app:cross-comparison}. The identity branch maps each target to $\tilde{\mathbf{t}}_{\text{id}} \in \mathbb{R}^D$, while the similarity branch outputs $\mathbf{c}\in\mathbb{R}^{D_u}$. The framework permits other target-only transformations, but every headline configuration uses a single DCN-v2 target self-cross. The component study varies this choice explicitly (Sec.~\ref{sec:ablation-summary} and Supp.~Sec.~\ref{sec:ablations}). A final MLP scores $[\mathbf{c};\tilde{\mathbf{t}}_{\text{id}}]$.

\noindent\textbf{Stage 1: Embedding and normalisation.}\label{sec:embedding}
Frozen pretrained item embeddings are projected to dimension $D$ and passed through shared LN for history and target, producing $\tilde{\mathbf{x}}_i, \tilde{\mathbf{t}}_c \in \mathbb{R}^D$. Only the history receives positional encodings. The shared normalization places query and key on a consistent scale for dot-product attention, since the frozen embeddings have heterogeneous magnitudes.

\noindent\textbf{Stage 2: Multi-head dot-product cross-attention.}\label{sec:cross-attn}
Each candidate attends to the user's history through multi-head cross-attention with query, key, and value projections. These projections map the Stage 1 width $D$ to $H$ heads of width $d$, producing per-head pooled histories $\mathbf{u}_h \in \mathbb{R}^d$ and a concatenated width $D_u=Hd$. The formulation permits $D_u\neq D$. This stage replaces DIN's MLP activation unit, which materialises a $[B, S, C, 4D]$ pairwise tensor, with batched matrix multiplications. Secs.~\ref{sec:public-evaluation} and~\ref{sec:production} measure the resulting latency, and Supp.~Sec.~\ref{app:flops} reports analytical FLOPs.

\noindent\textbf{Stage 3: Per-head gated-cross similarity.}\label{sec:per-head-cross}
The concatenated pooled history $\mathbf{u}\in\mathbb{R}^{D_u}$ is combined with a target projection $\mathbf{v}_t=\mathbf{W}_v\tilde{\mathbf{t}}\in\mathbb{R}^{D_u}$ through a DCN-v2-style operator that is linear in $\mathbf{u}$ for each fixed target. Relative to standard DCN-v2~\citep{wang2021dcn}, the operator uses two distinct inputs rather than self-crossing one vector, shares $K$ low-rank $d$-dimensional experts across heads, assigns each head a separate softmax gate computed from the full $\tilde{\mathbf{t}}\in\mathbb{R}^D$, and omits the residual. The expert matrices therefore contain $2Kdr$ parameters, while the gates contain $HKD$. Supp.~Sec.~\ref{app:cross-comparison} compares this count with DCN-v2 under general and equal-width settings.

The output $\mathbf{c} \in \mathbb{R}^{D_u}$ realises $\phi(\mathbf{u}, \tilde{\mathbf{t}}) = \mathbf{u} \odot \boldsymbol{\eta}(\tilde{\mathbf{t}})$ with a block-diagonal target map. Softmax gating and the expert mixture make this map non-linear in the target, but not in $\mathbf{u}$, so the PS equivalence is preserved. One layer retains this property under the evaluated recurrence. Omitting its residual keeps $\mathbf{c}$ as a pure interaction signal, although a $+\mathbf{u}$ residual would also remain history-linear. Deeper feature-interaction modules may follow $[\mathbf{c};\tilde{\mathbf{t}}_{\text{id}}]$ in the scorer without changing the equivalence. Supp.~Sec.~\ref{app:cross-comparison} gives the equations and a side-by-side comparison with DCN-v2.

\begin{remark}[Inference efficiency]
\label{rem:efficiency}
Under history-linearity, PS reduces similarity-operator calls from $S$ to one per candidate. \M{}'s cross-attention also permits request-local history reuse. A candidate-flattened evaluation duplicates the shared history across the $C$ candidate rows and re-encodes it for each row. The reused path instead computes the embedding, projection, positional encoding, LN, $\mathbf{K}$, and $\mathbf{V}$ once, expands the resulting tensors across the candidates, and discards them after the request. The public and industrial benchmarks compare these two paths within one request (Secs.~\ref{sec:public-evaluation} and~\ref{sec:production}), and Supp.~Sec.~\ref{app:efficiency} explains why reuse follows assembly of the positioned history and why a target-appending Transformer cannot reuse it.
\end{remark}

\begin{remark}[Amortisation across candidates]
\label{rem:amortisation}
Under request-local reuse, the per-request computation separates into history-only work, performed once, and target-dependent work, performed for each candidate. Let their costs be $T_{\text{shared}}$ and $T_{\text{per-cand}}$, respectively. Candidate-flattened evaluation performs $C(T_{\text{shared}}+T_{\text{per-cand}})$ work, whereas reuse performs $T_{\text{shared}}+CT_{\text{per-cand}}$. Its amortised per-candidate cost is $T_{\text{shared}}/C+T_{\text{per-cand}}$. As $C$ grows, this cost approaches $T_{\text{per-cand}}$. On Industrial-B at $C=1{,}000$, shared history work accounts for about $3\%$ of the reused-path latency.
\end{remark}

\section{Experiments}
\label{sec:experiments}

We organise the evaluation around four research questions:
\begin{itemize}[wide=0pt,nosep]
\item \textbf{RQ1: Public ranking quality.} Does the fixed, prespecified \M{} configuration improve ranking metrics over target-aware and target-agnostic baselines?
\item \textbf{RQ2: Candidate-set scaling.} How do candidate count and request-local history reuse affect latency, as anticipated by Remarks~\ref{rem:efficiency} and~\ref{rem:amortisation}?
\item \textbf{RQ3: Industrial transfer.} Does the public quality--latency direction persist on logged industrial requests?
\item \textbf{RQ4: Component evidence.} Which component comparisons support the fixed architectural defaults?
\end{itemize}
The main text specifies the comparison design needed to interpret the results, while Supp.~Secs.~\ref{app:experimental-details} and~\ref{app:latency-detail} provide the architectural, optimisation, and timing details required for reproduction.

\subsection{Public Dataset Evaluation}
\label{sec:public-evaluation}

\noindent\textbf{Datasets and protocol.} We use Steam~\citep{kang2018sasrec} (281{,}354 users, 11{,}961 items) and the 2019 MovieLens-25M (\mbox{ML-25M}) release~\citep{grouplens2019movielens25m} (162{,}541 users, 32{,}720 items) under the same per-user leave-one-out protocol. We retain all Steam prefixes, yielding 2{,}706{,}396 training examples, and the $20$ most recent \mbox{ML-25M} prefixes per user, yielding~3{,}226{,}917. Each test item is ranked against~$99$ unique negatives that are disjoint from the user's full sequence and identical across models. This sampled protocol is common, but sampled metrics can order models differently from full-catalogue ranking~\citep{krichene2020sampled}. We therefore also re-score one training run per model, without retraining, against $999$ negatives retrieved by MiniLM nearest-neighbour search and against the full catalogue, counting ties against the test item (Tab.~\ref{tab:robust}). Details are in Supp.~Sec.~\ref{app:experimental-details}.

\noindent\textbf{Model inputs.} All models share frozen 384-dimensional item-text embeddings (Supp.~Sec.~\ref{app:experimental-details}), a trainable projection to $D{=}64$, and histories truncated to at most $50$ items.

\noindent\textbf{Baseline embedding and Transformer models.} The main baselines share the embedding projection, learned positions, and LN. \textbf{SASRec}~\citep{kang2018sasrec} is target-agnostic. A two-layer causal self-attention Transformer encodes the history into a fixed vector, and scoring uses the dot product $\mathbf{u}^\top \tilde{\mathbf{t}}$. The \emph{target-aware Transformer}~(TA-Transformer) baseline appends each candidate to the history, applies one layer of bidirectional self-attention with one 64-dimensional head, and scores the candidate-position output. We prespecify this depth and head count to match the industrial comparator. Supp.~Sec.~\ref{app:experimental-details} documents its public-to-industrial adaptations.

\noindent\textbf{DIN baselines and target normalisation.} \textbf{DIN}~\citep{zhou2018deep} uses a target-aware MLP activation unit to compute per-item attention weights, pools the history into a single vector $\mathbf{u}$, and scores via $\mathrm{MLP}([\mathbf{u}, \mathbf{t}])$. \textbf{DIN + DCN-v2} augments DIN with a DCN-v2 bilinear cross network on $(\mathbf{u}, \mathbf{t})$. Writing its output as $\mathbf{c}_{\mathrm{DCN}}$, the model scores via $\mathrm{MLP}([\mathbf{u}, \mathbf{t}, \mathbf{c}_{\mathrm{DCN}}])$. The headline SASRec and both DIN variants apply the shared LN to the target. Tab.~\ref{tab:main-full} also reports SASRec without this target normalisation. We also evaluate CAN~\citep{bian2022can}, whose target-parameterised MLP acts on each history item, and DIEN~\citep{zhou2019dien}, whose target-conditioned gated recurrent unit models interest evolution (Tab.~\ref{tab:robust}). On Steam at $C=1{,}000$, their request-local latencies are about $2.0$ and $16.4$\,ms.

\noindent\textbf{Fixed \M{} configuration.} Our prespecified configuration uses one residual-free cross layer with $4$ experts, a DCN-v2 target self-cross, and an MLP that receives only the crossed outputs. We use it on both datasets without per-dataset selection.

\noindent\textbf{SASRec target normalisation.} Applying the shared LN to the target raises SASRec's \emph{mean reciprocal rank}~(MRR) by $15.4\%$ on Steam ($0.3736 \rightarrow 0.4311$) and $41.6\%$ on ML-25M ($0.4658 \rightarrow 0.6597$). Without target normalisation, the dot product has a scale mismatch unrelated to target-agnosticism. This correction is larger than most measured architectural effects, so we use the normalised variant for the main comparison and retain both variants in Tab.~\ref{tab:main-full}.

\noindent\textbf{Training, model selection, and metrics.} All models use \emph{Bayesian Personalized Ranking}~(BPR)~\citep{rendle2009bpr}, and checkpoint selection uses validation MRR. We report \emph{hit rate}~(HIT)@$\{1,5,10\}$, NDCG@$\{5,10\}$, and MRR. Tab.~\ref{tab:main} averages three training seeds; other main-text ranking results use one training run per model. Supp.~Sec.~\ref{app:experimental-details} gives optimisation, batching, and model-size details (Tab.~\ref{tab:params}).

\begin{table*}[t]
\centering
\small
\caption{Public-dataset results (HIT@10, NDCG@10, MRR): mean $\pm$ sample standard deviation over three training seeds. \M{} is the prespecified single-layer, $4$-expert configuration of Sec.~\ref{sec:architecture}, identical on both datasets and not selected per dataset. Tab.~\ref{tab:main-full} reports all combinations of $\{4,8\}$ experts and $\{1,2\}$ layers from one training run each. SASRec uses the normalised target.}
\label{tab:main}
\setlength{\tabcolsep}{2.5pt}
\begin{tabular}{@{}l ccc ccc@{}}
\toprule
& \multicolumn{3}{c}{Steam} & \multicolumn{3}{c}{ML-25M} \\
\cmidrule(lr){2-4} \cmidrule(lr){5-7}
Model & HIT@10 & NDCG@10 & MRR & HIT@10 & NDCG@10 & MRR \\
\midrule
SASRec (target-agnostic) & .7677\,{\scriptsize$\pm$.0032} & .4997\,{\scriptsize$\pm$.0029} & .4279\,{\scriptsize$\pm$.0029} & .9327\,{\scriptsize$\pm$.0017} & .7234\,{\scriptsize$\pm$.0038} & .6595\,{\scriptsize$\pm$.0043} \\
TA-Transformer (1 layer, 1 head) & .7741\,{\scriptsize$\pm$.0043} & .5128\,{\scriptsize$\pm$.0062} & .4425\,{\scriptsize$\pm$.0065} & .9283\,{\scriptsize$\pm$.0016} & .7158\,{\scriptsize$\pm$.0031} & .6512\,{\scriptsize$\pm$.0035} \\
DIN & .7817\,{\scriptsize$\pm$.0029} & .5253\,{\scriptsize$\pm$.0045} & .4561\,{\scriptsize$\pm$.0049} & .9399\,{\scriptsize$\pm$.0021} & .7347\,{\scriptsize$\pm$.0063} & .6719\,{\scriptsize$\pm$.0074} \\
DIN + DCN-v2 & .7851\,{\scriptsize$\pm$.0029} & .5300\,{\scriptsize$\pm$.0021} & .4609\,{\scriptsize$\pm$.0017} & .9436\,{\scriptsize$\pm$.0003} & .7467\,{\scriptsize$\pm$.0022} & .6862\,{\scriptsize$\pm$.0029} \\
\M{} (4 experts, 1 layer) & \textbf{.7940}\,{\scriptsize$\pm$.0019} & \textbf{.5409}\,{\scriptsize$\pm$.0016} & \textbf{.4720}\,{\scriptsize$\pm$.0016} & \textbf{.9503}\,{\scriptsize$\pm$.0006} & \textbf{.7605}\,{\scriptsize$\pm$.0016} & \textbf{.7017}\,{\scriptsize$\pm$.0019} \\
\bottomrule
\end{tabular}
\end{table*}
\begin{table*}[t]
\centering
\small
\caption{Full public-dataset results from one training run per model. Bold marks best per dataset and metric. ``SASRec (history-only LN)'' omits the shared LN on the target, while the main SASRec row applies it. Supp.~Sec.~\ref{app:experimental-details} gives batching details.}
\label{tab:main-full}
\setlength{\tabcolsep}{3pt}
\begin{tabular}{@{}llcccccc@{}}
\toprule
Dataset & Model & HIT@1 & HIT@5 & HIT@10 & NDCG@5 & NDCG@10 & MRR \\
\midrule
\multirow{9}{*}{Steam}
& SASRec (history-only LN) & 0.2168 & 0.5554 & 0.7198 & 0.3917 & 0.4451 & 0.3736 \\
& SASRec & 0.2738 & 0.6198 & 0.7704 & 0.4540 & 0.5029 & 0.4311 \\
& TA-Transformer (1 layer, 1 head) & 0.2809 & 0.6235 & 0.7695 & 0.4599 & 0.5073 & 0.4370 \\
& DIN & 0.3082 & 0.6465 & 0.7848 & 0.4855 & 0.5304 & 0.4616 \\
& DIN + DCN-v2 & 0.3060 & 0.6439 & 0.7823 & 0.4834 & 0.5284 & 0.4598 \\
\cmidrule(lr){2-8}
& \M{} (4 experts, 1 layer) & 0.3196 & 0.6563 & 0.7923 & 0.4963 & 0.5405 & 0.4720 \\
& \M{} (4 experts, 2 layers) & 0.3238 & 0.6628 & 0.7960 & 0.5022 & 0.5455 & 0.4771 \\
& \M{} (8 experts, 1 layer) & \textbf{0.3272} & 0.6678 & 0.8000 & \textbf{0.5064} & \textbf{0.5494} & \textbf{0.4808} \\
& \M{} (8 experts, 2 layers) & 0.3257 & \textbf{0.6680} & \textbf{0.8019} & 0.5058 & 0.5493 & 0.4801 \\
\midrule
\multirow{9}{*}{ML-25M}
& SASRec (history-only LN) & 0.3140 & 0.6472 & 0.7714 & 0.4901 & 0.5304 & 0.4658 \\
& SASRec & 0.5084 & 0.8587 & 0.9333 & 0.6992 & 0.7236 & 0.6597 \\
& TA-Transformer (1 layer, 1 head) & 0.5027 & 0.8545 & 0.9302 & 0.6942 & 0.7189 & 0.6546 \\
& DIN & 0.5295 & 0.8700 & 0.9408 & 0.7151 & 0.7383 & 0.6762 \\
& DIN + DCN-v2 & 0.5405 & 0.8753 & 0.9433 & 0.7234 & 0.7457 & 0.6850 \\
\cmidrule(lr){2-8}
& \M{} (4 experts, 1 layer) & 0.5634 & 0.8879 & 0.9510 & 0.7412 & 0.7619 & 0.7033 \\
& \M{} (4 experts, 2 layers) & 0.5698 & 0.8913 & 0.9531 & 0.7461 & 0.7664 & 0.7086 \\
& \M{} (8 experts, 1 layer) & 0.5686 & 0.8912 & 0.9525 & 0.7458 & 0.7660 & 0.7081 \\
& \M{} (8 experts, 2 layers) & \textbf{0.5745} & \textbf{0.8949} & \textbf{0.9548} & \textbf{0.7505} & \textbf{0.7701} & \textbf{0.7128} \\
\bottomrule
\end{tabular}
\end{table*}
\noindent\textbf{RQ1: Public ranking quality.} Over three training seeds (Tab.~\ref{tab:main}), the prespecified \M{} configuration records the highest mean on every metric on both datasets, and among these five models it is highest on all six metrics of Tab.~\ref{tab:main-full} in every seed. Its mean MRR gain over DIN is $+3.5\%$ on Steam and $+4.4\%$ on ML-25M. Across the three metrics of Tab.~\ref{tab:main}, its margins over the strongest baseline span Steam~($0.0089$--$0.0111$) and ML-25M~($0.0067$--$0.0155$). For one training run per model, paired-bootstrap 95\% intervals over test users (10{,}000 resamples) of \M{} minus each baseline exclude zero for HIT@10, NDCG@10, and MRR on both datasets. The TA-Transformer is a competitive mid-table target-aware baseline: its mean MRR is $0.0146$ above normalised SASRec on Steam and $0.0083$ below it on ML-25M, while it remains below both DIN variants and \M{} on every metric. Relative to TA-Transformer, \M{} raises mean MRR by~$0.0295$ on Steam and $0.0505$ on ML-25M. A parameter-matched TA-Transformer has higher values on all six Steam sampled metrics but is $16\times$ slower at $C=1{,}000$ (Supp.~Sec.~\ref{app:capacity-ta}).

\noindent\textbf{Consistency across the tested configurations.} Across the four combinations of $\{4,8\}$ experts and $\{1,2\}$ layers (Tab.~\ref{tab:main-full}), all model variants exceed the strongest baseline in MRR by $0.0104$ to~$0.0192$ on Steam and by $0.0183$ to~$0.0278$ on ML-25M. All four also exceed it in HIT@10 and NDCG@10 on both datasets. Selecting by validation MRR would choose the $8$-expert, $2$-layer variant on both datasets. That variant has the highest test MRR on ML-25M. On Steam, its test MRR of~$0.4801$ is $0.0007$ below the $8$-expert, $1$-layer variant.
\begin{table}[t]
\centering
\scriptsize
\caption{Single-run MRR under sampled ($99$ uniform negatives), hard ($999$ negatives from a fixed MiniLM retriever), and full-catalogue ranking (the test item versus every other item outside the user's sequence).}
\label{tab:robust}
\begin{tabular}{@{}lrrr@{}}
\toprule
Model & Sampled & Hard & Full \\
\midrule
\multicolumn{4}{c}{Steam} \\
SASRec & .4311 & .0636 & .0351 \\ %
TA-Transformer (1L, 1H) & .4370 & .0623 & .0340 \\ %
DIN & .4616 & .0748 & .0457 \\ %
DIN + DCN-v2 & .4598 & .0731 & .0434 \\ %
CAN & .4611 & .0785 & .0388 \\ %
DIEN & .4495 & .0703 & .0380 \\ %
\M{} (4 experts, 1 layer) & \textbf{.4720} & \textbf{.0804} & \textbf{.0466} \\ %
\midrule
\multicolumn{4}{c}{ML-25M} \\
SASRec & .6597 & .0839 & .0374 \\ %
TA-Transformer (1L, 1H) & .6546 & .0942 & .0383 \\ %
DIN & .6762 & .0947 & .0471 \\ %
DIN + DCN-v2 & .6850 & .1061 & .0485 \\ %
CAN & .6630 & .0940 & .0378 \\ %
DIEN & .6769 & .0975 & .0415 \\ %
\M{} (4 experts, 1 layer) & \textbf{.7033} & \textbf{.1099} & \textbf{.0565} \\ %
\bottomrule
\end{tabular}
\end{table}

\noindent\textbf{Harder candidate sets.} Among the models in Tab.~\ref{tab:robust}, \M{} has the highest MRR with hard negatives on both datasets, and its smallest margin is $+0.0019$ over CAN on Steam (Tab.~\ref{tab:robust}).
Under full-catalogue ranking, its MRR is higher by $0.0009$ than DIN's on Steam and by $0.0079$ than that of DIN + DCN-v2 on ML-25M, although DIN has higher Steam HIT@1 ($0.0186$ versus $0.0177$).
CAN and DIEN remain below \M{} under all three protocols on both datasets, and every paired-bootstrap $95\%$ MRR interval for \M{} minus a listed baseline is above zero.
The SASRec and TA-Transformer MRR order reverses between sampled and full ranking on both datasets, consistent with~\citet{krichene2020sampled}.

\noindent\textbf{\mbox{RQ2: Candidate-set} scaling and cacheability.} We measure scoring latency across candidate-set sizes with candidate-flattened history re-encoding and request-local history reuse, testing the serving mechanisms in Remarks~\ref{rem:efficiency} and~\ref{rem:amortisation}. Supp.~Sec.~\ref{app:flops} provides complementary analytical FLOP estimates.

\noindent\textbf{Measurement protocol.} We time one user, one $S=50$ history, and $C$ candidates sharing it on a dedicated processor instance with 32 virtual cores. For every model, each replicate measures candidate-flattened re-encoding and request-local reuse in the same process. No state persists across calls. We report medians across five process-isolated replicate minima and median paired speedups. Supp.~Sec.~\ref{app:latency-detail} gives the software environment, timing procedure, execution order, variability, and output-equivalence check.

\begin{table*}[t]
\centering
\scriptsize
\caption{Public quality--latency comparison. Top: NDCG@10, candidate-flattened latency (Flat., ms), request-local-reuse latency (Reuse, ms), and median paired reuse benefit at $C=1{,}000$. NDCG@10 is from one training run, and $\Delta$ is its difference from TA-Transformer. Bottom: request-local-reuse latency at smaller $C$. Latencies are medians of five replicate minima. Benefit is the median within-replicate Flat./Reuse ratio, not the quotient of the displayed medians. TA-Trans.\ is TA-Transformer and DIN+DCN is DIN + DCN-v2.}
\label{tab:latency-public}
\setlength{\tabcolsep}{1.8pt}
\begin{tabular}{@{}lrrrrrrrrrr@{}}
\toprule
& \multicolumn{5}{c}{Steam} & \multicolumn{5}{c}{ML-25M} \\
\cmidrule(lr){2-6} \cmidrule(lr){7-11}
Model & NDCG@10 & $\Delta$ & Flat. & Reuse & Benefit & NDCG@10 & $\Delta$ & Flat. & Reuse & Benefit \\
\midrule
TA-Transformer (1 layer, 1 head) & $0.5073$ & $0.0000$ & $14.76$ & $15.09$ & $1.03\times$ & $0.7189$ & $0.0000$ & $14.52$ & $13.74$ & $1.05\times$ \\
DIN & $0.5304$ & $+0.0231$ & $16.57$ & $9.89$ & $1.67\times$ & $0.7383$ & $+0.0194$ & $18.70$ & $9.87$ & $1.89\times$ \\
DIN + DCN-v2 & $0.5284$ & $+0.0211$ & $17.74$ & $10.58$ & $1.80\times$ & $0.7457$ & $+0.0268$ & $16.64$ & $10.19$ & $1.66\times$ \\
\M{} (4 experts, 1 layer) & $0.5405$ & $+0.0332$ & $27.86$ & $1.97$ & $14.21\times$ & $0.7619$ & $+0.0430$ & $29.30$ & $2.00$ & $14.68\times$ \\
\bottomrule
\end{tabular}
\vspace{2pt}

\begin{tabular}{@{}lrrrrrrrr@{}}
\toprule
& \multicolumn{4}{c}{Steam} & \multicolumn{4}{c}{ML-25M} \\
\cmidrule(lr){2-5} \cmidrule(lr){6-9}
$C$ & TA-Trans. & DIN & DIN+DCN & \M{} & TA-Trans. & DIN & DIN+DCN & \M{} \\
\midrule
$1$ & $0.60$ & $0.36$ & $0.67$ & $0.80$ & $0.60$ & $0.35$ & $0.67$ & $0.80$ \\
$10$ & $0.87$ & $0.51$ & $0.82$ & $0.85$ & $0.87$ & $0.51$ & $0.82$ & $0.85$ \\
$100$ & $1.50$ & $0.63$ & $1.07$ & $1.21$ & $1.49$ & $0.63$ & $1.07$ & $1.21$ \\
\bottomrule
\end{tabular}
\end{table*}

\begin{figure*}[t]
\centering
\includegraphics{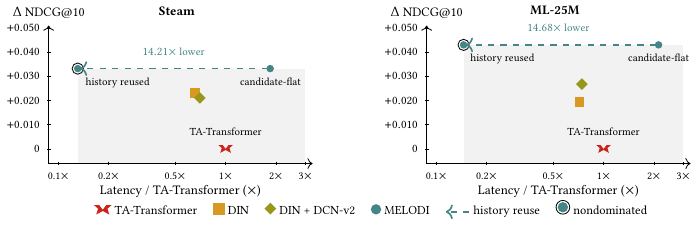}
\caption{Public quality gain versus latency relative to TA-Transformer at $C=1{,}000$. Quality is the single-run NDCG@10 difference from TA-Transformer. Baseline points use request-local-reuse latency over TA-Transformer's. \M{}'s dashed arrow shows request-local reuse lowering its latency, and upper-left points have higher NDCG@10 at lower latency. Grey regions are dominated by reused \M{}, and black rings mark the nondominated plotted operating points. Target-agnostic SASRec (NDCG@10 $0.5029$ on Steam and $0.7236$ on ML-25M) is omitted to focus on candidate-conditioned models.}
\Description{Two quality-versus-latency panels for Steam and ML-25M. Request-local-reuse MELODI lies above and to the left of TA-Transformer, DIN, and DIN plus DCN-v2. A dashed arrow from candidate-flattened to request-local-reuse MELODI shows the same-process reduction.}
\label{fig:public-pareto}
\end{figure*}

\noindent\textbf{RQ1--RQ2 quality--latency comparison.} Fig.~\ref{fig:public-pareto} is the public counterpart to Supp.~Fig.~\ref{fig:pareto}, with TA-Transformer as the target-aware reference. The prespecified \M{} configuration has higher NDCG@10 than TA-Transformer by $0.0332$ on Steam and $0.0430$ on ML-25M. At $C=1{,}000$ with request-local reuse, it records $1.97$ and $2.00$\,ms, against $15.09$ and $13.74$\,ms for TA-Transformer and $9.89$ and $9.87$\,ms for DIN. \M{} is faster than TA-Transformer by $7.7\times$ and $6.9\times$, and than DIN by $5.0\times$ and $4.9\times$. Baseline latencies at this $C$ vary by 9--25\% between runs, and a repeat run gives $8.6$--$8.7\times$ and $3.8$--$4.0\times$ (Supp.~Sec.~\ref{app:latency-detail}). Its latency is nearly equal across datasets because they share $S$, $D$, and the architecture, and embedding-table size does not affect per-request cost.

\noindent\textbf{Candidate-set boundary and deployment regime.} Under request-local reuse, DIN is fastest at $C\in\{1,10,100\}$ in both datasets, where \M{} is $2.2$--$2.3\times$, $1.7\times$, and $1.9\times$ slower. \M{} has a fixed history cost before candidate scoring, whereas DIN's activation unit vectorises efficiently until its $[B,S,C,4D]$ tensor becomes large. From $C=100$ to $1{,}000$, DIN's latency rises from $0.63$ to about $9.9$\,ms and \M{}'s from $1.21$ to about $2.0$\,ms, so their order reverses at a hardware-dependent point in between, which a finer sweep places between $C=500$ and $700$, and between $200$ and $300$ for DIN + DCN-v2 (Supp.~Sec.~\ref{app:latency-detail}). On this processor, a ranker that scores about $100$ candidates per request therefore trades \M{}'s higher NDCG@10 ($+0.0101$ on Steam, $+0.0236$ on ML-25M) against DIN's $1.9\times$ lower latency, whereas at about $1{,}000$ candidates \M{} offers both higher NDCG@10 and lower latency. The $C=1$ regime also explains why \M{} can be slower in a training loop that scores one target at a time.

\noindent\textbf{Request-local history reuse.} At $C=1{,}000$, request-local reuse lowers \M{} latency on Steam from $27.86$ to $1.97$\,ms and on ML-25M from $29.30$ to $2.00$\,ms. The median paired benefit is $14.21\times$ on Steam and $14.68\times$ on ML-25M (replicate ranges in Supp.~Sec.~\ref{app:latency-detail}). It grows from $1.00\times$ at $C=1$ to $1.17\times$ at $C=10$ and $1.43$--$1.45\times$ at $C=100$. Reuse changes TA-Transformer latency by only $1.03$--$1.05\times$ because self-attention mixes each candidate with its history. Sharing the target-independent input path gives DIN a $1.67$--$1.89\times$ benefit and DIN + DCN-v2 a $1.66$--$1.80\times$ benefit, although their activation units remain candidate-dependent.

\subsection{Industrial Dataset Evaluation}
\label{sec:production}
\setcounter{dbltopnumber}{3}
\renewcommand{\dbltopfraction}{0.9}

\noindent\textbf{Datasets and protocol.} We evaluate two large-scale industrial ranking datasets, denoted \emph{Industrial-A} and \emph{Industrial-B}, that consist of logged requests. Both are proprietary and cannot be publicly released. The Industrial-A test set contains 591{,}286 candidate rows from~5{,}836 requests. The Industrial-B test set contains~1{,}088{,}853 candidate rows from~5{,}625 requests. Every evaluated candidate has already passed upstream candidate generation, so the binary purchase task uses hard negatives from the displayed candidate set. Supp.~Sec.~\ref{app:experimental-details} gives the temporal splits and reported segments.

\noindent\textbf{Model inputs.} For each dataset, all models receive the same request-level features and item representations. Sequential models also receive the same two event-history partitions.

\noindent\textbf{Baselines and configurations.} Each dataset has its own tabular baseline, a TA-Transformer, and DIN. Both sequential baselines use one branch for each of the same two event-history partitions. \M{} uses one similarity cross layer and a DCN-v2 target self-cross in both datasets, but the two datasets use separately configured widths, expert counts, and event-history crosses. Supp.~Sec.~\ref{app:experimental-details} gives the exact configurations.

\noindent\textbf{Training, model selection, and metrics.} All models use binary cross-entropy and are selected by validation NDCG@20. We report four metrics: \emph{area under the receiver operating characteristic curve}~(AUROC), AUPRC, NDCG@20, and Recall@20. We report each metric as a percentage change relative to the tabular baseline, computed from three-decimal values.

\begin{table*}[t]
\centering
\small
\caption{Industrial-A: \% change from the tabular baseline, overall and by segment (N@20: NDCG@20, R@20: Recall@20).}
\label{tab:production-a}
\setlength{\tabcolsep}{2.5pt}
\begin{tabular}{@{}ll cccc cccc cccc@{}}
\toprule
& & \multicolumn{4}{c}{Overall} & \multicolumn{4}{c}{Segment-1} & \multicolumn{4}{c}{Segment-2} \\
\cmidrule(lr){3-6} \cmidrule(lr){7-10} \cmidrule(lr){11-14}
& Model & AUROC & AUPRC & N@20 & R@20 & AUROC & AUPRC & N@20 & R@20 & AUROC & AUPRC & N@20 & R@20 \\
\midrule
& Tabular baseline & -- & -- & -- & -- & -- & -- & -- & -- & -- & -- & -- & -- \\
& TA-Transformer & +2.5 & +22.2 & +8.8 & +5.8 & +2.3 & +28.6 & +11.1 & +3.8 & +3.1 & \textbf{+11.1} & +6.9 & +7.1 \\
& DIN & +2.8 & +16.7 & +8.8 & +6.0 & \textbf{+2.6} & +22.4 & +11.8 & +4.6 & +3.4 & +7.4 & +6.4 & +6.7 \\
& \M{} & \textbf{+2.9} & \textbf{+25.0} & \textbf{+11.3} & \textbf{+7.1} & +2.5 & \textbf{+32.7} & \textbf{+15.4} & \textbf{+5.5} & \textbf{+3.6} & \textbf{+11.1} & \textbf{+7.9} & \textbf{+8.1} \\
\bottomrule
\end{tabular}
\end{table*}
\begin{table*}[t]
\centering
\small
\caption{Industrial-B: \% change from the tabular baseline, overall and by identifier availability, as in Tab.~\ref{tab:production-a}.}
\label{tab:production-b}
\setlength{\tabcolsep}{2.5pt}
\begin{tabular}{@{}ll cccc cccc cccc@{}}
\toprule
& & \multicolumn{4}{c}{Overall} & \multicolumn{4}{c}{Identifier available} & \multicolumn{4}{c}{Identifier unavailable} \\
\cmidrule(lr){3-6} \cmidrule(lr){7-10} \cmidrule(lr){11-14}
& Model & AUROC & AUPRC & N@20 & R@20 & AUROC & AUPRC & N@20 & R@20 & AUROC & AUPRC & N@20 & R@20 \\
\midrule
& Tabular baseline & -- & -- & -- & -- & -- & -- & -- & -- & -- & -- & -- & -- \\
& TA-Transformer & +2.0 & +6.5 & +2.3 & +2.3 & +2.3 & $-$3.3 & +3.5 & +4.9 & +1.6 & \textbf{+12.5} & \textbf{+2.3} & +0.9 \\
& DIN & \textbf{+2.5} & +6.5 & +2.7 & +3.1 & \textbf{+3.1} & \textbf{+3.3} & \textbf{+5.6} & +7.4 & \textbf{+1.9} & +10.2 & +1.6 & 0.0 \\
& \M{} & +2.0 & \textbf{+7.8} & \textbf{+3.2} & \textbf{+4.2} & +2.8 & \textbf{+3.3} & \textbf{+5.6} & \textbf{+8.4} & +1.0 & +10.2 & \textbf{+2.3} & \textbf{+1.3} \\
\bottomrule
\end{tabular}
\end{table*}
\noindent\textbf{RQ3: Industrial transfer.} \textbf{Industrial-A.} Overall (Tab.~\ref{tab:production-a}), \M{} is highest on all four metrics. Relative to the tabular baseline, it raises AUPRC by $25.0\%$ and NDCG@20 by $11.3\%$. In both segments, it records the highest NDCG@20, Recall@20, and AUPRC, tying TA-Transformer on Segment-2 AUPRC at reported precision, while DIN has a higher Segment-1 AUROC ($+2.6\%$ vs.\ $+2.5\%$).

\noindent\textbf{Industrial-B.} Overall (Tab.~\ref{tab:production-b}), \M{} records the highest AUPRC, NDCG@20, and Recall@20, while DIN has the highest AUROC ($+2.5\%$ vs.\ $+2.0\%$). Relative to the tabular baseline, \M{} raises AUPRC by $7.8\%$, NDCG@20 by $3.2\%$, and Recall@20 by $4.2\%$. With and without a user identifier, it records the highest Recall@20 and ties the highest NDCG@20 at reported precision. It also ties the highest AUPRC when the identifier is available.

\noindent\textbf{Latency and cacheability.} We time one user with $50$ history items and $1{,}000$ candidates on the same commodity processor instance with 32 virtual cores used for the public benchmarks. Supp.~Sec.~\ref{app:latency-detail} gives the timing procedure and latencies relative to each tabular baseline for the candidate-flattened path, which duplicates the history across candidates, and the request-local-reuse path, which encodes one shared history for all candidates in the request.

\noindent\textbf{Quality--latency results.} Supp.~Fig.~\ref{fig:pareto} plots NDCG@20 gain against latency overhead relative to each tabular baseline. \M{} increases NDCG@20 by $11.3\%$ on Industrial-A and $3.2\%$ on Industrial-B. In both datasets, it has higher reported NDCG@20 than TA-Transformer and DIN, and its candidate-flattened path is $5.0$--$7.6\times$ and $2.0$--$2.9\times$ faster, respectively.

\noindent\textbf{Request-local history reuse.} Encoding the history once per request reduces \M{}'s latency by a further $2.6$--$3.8\times$, to $13.0$--$28.7\times$ lower than TA-Transformer and $5.2$--$11.1\times$ lower than DIN.
Both benchmarks time the same two paths, so the benefit difference reflects how much computation remains candidate-dependent. In the public $D{=}64$ configuration, the target-dependent path is light and repeated history processing dominates the candidate-flattened cost, so the $16.02\times$ analytical FLOP ratio at $C{=}1{,}000$ is close to the measured \mbox{$14.21$--$14.68\times$} benefit. The industrial configurations' $D{=}256$, tabular features, and wider similarity outputs and scorers leave more candidate-dependent computation after reuse, which explains the smaller measured benefit.

\subsection{Component-Ablation Evidence}
\label{sec:ablation-summary}

\begin{table}[t]
\centering
\scriptsize
\caption{Component ablations (MRR, 4 experts). Rows 3 and 4 include the coupled and isolated interventions. Supp.~Sec.~\ref{sec:ablations} gives deltas and details.}
\label{tab:ablations}
\setlength{\tabcolsep}{2.5pt}
\begin{tabular}{@{}r p{5.2cm} cc@{}}
\toprule
\# & Configuration & Steam & ML-25M \\
\midrule
1 & \textbf{Anchor: 1 layer, no residual, cross-only scorer} & \textbf{.4720} & \textbf{.7033} \\
2 & $-$ target self-cross & .4702 & .6882 \\
3a & $+$ target residual & .4722 & .7023 \\
3b & $+$ pooled-history residual & .4716 & .7023 \\
4a & $+$ second layer; identity also deepened & .4771 & .7086 \\
4b & $+$ second layer; identity fixed & .4794 & .7019 \\
5 & $+$ uncrossed pooled history and projected target & .4758 & .7038 \\
6 & $+$ projected target only & .4643 & .6934 \\
7 & $-$ similarity cross; $\mathbf{u}$ to the scorer & .4649 & .6924 \\
\bottomrule
\end{tabular}
\end{table}

\noindent\textbf{RQ4: Component evidence.} Tab.~\ref{tab:ablations} compares component changes with the anchor configuration. Supp.~Sec.~\ref{sec:ablations} specifies the isolated and coupled interventions. Two observed patterns motivate the fixed defaults. Removing the tested target self-cross identity branch lowers MRR by $0.0018$ on Steam and $0.0151$ on ML-25M. Adding the projected target alone lowers MRR by $0.0077$ on Steam and $0.0099$ on ML-25M. When the projected target is supplied, adding the corresponding uncrossed pooled history (row 5 rather than row 6) yields the higher value on both. Removing the similarity cross itself (row 7) lowers MRR by $0.0071$ on Steam and $0.0109$ on ML-25M and saves about $0.8$\,ms at $C=1{,}000$, although it raises MRR by $0.0011$ with hard negatives on ML-25M.

The two residual variants change MRR by at most $0.0010$. A second similarity layer with identity capacity fixed changes MRR by $+0.0074$ on Steam and $-0.0014$ on ML-25M, so its direction is dataset-dependent. Expert-count changes likewise differ across settings (Tab.~\ref{tab:experts}). We retain one layer so that the stated SP/PS guarantee applies, and omit its residual so that the branch outputs only the interaction between history and target. Each configuration is represented by one training run, so MRR differences of a few thousandths are not established.

\section{Conclusion}

We proposed a \emph{similarity--identity decoupling} framework for target-aware ranking. Proposition~\ref{prop:sp-ps-equivalence} gives the linearity condition under which the cheaper PS order exactly recovers SP, and \M{} realises this mechanism with multi-head cross-attention and a residual-free target-dependent cross layer. Averaged over three training seeds, its fixed public configuration exceeds every baseline in Tab.~\ref{tab:main} on every reported metric. Request-local reuse gives it lower latency than the candidate-conditioned baselines at $C=1{,}000$, while DIN remains faster up to $C=500$. In the industrial evaluation, \M{} matches or exceeds the reported AUPRC, NDCG, and Recall of the target-aware baselines at lower latency. Single-run component comparisons motivate fixed defaults, chiefly the target self-cross and pairing the projected target with pooled history, but do not establish small effects. Limitations remain: the evaluation is offline, without an online A/B test, the industrial data cannot be released, and latency is measured on one processor type.

\section{Ethical Considerations}
\label{sec:ethics}

The industrial evaluation raises two concerns: bias in logged exposure and the handling of behavioural data.

\noindent\textbf{Exposure and position bias in the training signal.} The model is trained on logged purchase labels for displayed candidates, which reflect the exposure and position biases of the logging policy. \M{} does not correct for them, so it can inherit and perpetuate them. The framework is agnostic to the training loss, so corrections such as inverse-propensity weighting~\citep{schnabel2016recommendations} or a randomised-exposure holdout for estimating propensities need no architectural change. This bias limits the interpretation of the industrial results.

\noindent\textbf{Privacy and data handling.} Interaction histories are behavioural data and may be identifying even without explicit user identifiers. Request-local history tensors therefore need the same access controls as their source histories and should be discarded once the current candidate set has been scored.

\bibliographystyle{ACM-Reference-Format}
\bibliography{references}

@inproceedings{zhou2018deep,
  title={Deep Interest Network for Click-Through Rate Prediction},
  author={Zhou, Guorui and Zhu, Xiaoqiang and Song, Chenru and Fan, Ying and Zhu, Han and Ma, Xiao and Yan, Yanghui and Jin, Junqi and Li, Han and Gai, Kun},
  booktitle={Proceedings of the 24th ACM SIGKDD International Conference on Knowledge Discovery \& Data Mining},
  pages={1059--1068},
  year={2018},
  publisher={Association for Computing Machinery},
  address={New York, NY, USA}
}

@inproceedings{wang2021dcn,
  title={{DCN V2}: Improved Deep \& Cross Network and Practical Lessons for Web-scale Learning to Rank Systems},
  author={Wang, Ruoxi and Shivanna, Rakesh and Cheng, Derek and Jain, Sagar and Lin, Dong and Hong, Lichan and Chi, Ed},
  booktitle={Proceedings of the Web Conference 2021},
  pages={1785--1797},
  year={2021},
  publisher={Association for Computing Machinery},
  address={New York, NY, USA}
}

@inproceedings{chen2019bst,
  title={Behavior Sequence Transformer for E-commerce Recommendation in Alibaba},
  author={Chen, Qiwei and Zhao, Huan and Li, Wei and Huang, Pipei and Ou, Wenwu},
  booktitle={Proceedings of the 1st International Workshop on Deep Learning Practice for High-Dimensional Sparse Data},
  pages={1--4},
  year={2019},
  publisher={Association for Computing Machinery},
  address={New York, NY, USA}
}

@inproceedings{feng2025dare,
  title={Long-Sequence Recommendation Models Need Decoupled Embeddings},
  author={Feng, Ningya and Pan, Junwei and Wu, Jialong and Chen, Baixu and Wang, Ximei and Li, Qian and Hu, Xian and Jiang, Jie and Long, Mingsheng},
  booktitle={International Conference on Learning Representations (ICLR)},
  year={2025},
  volume={2025},
  pages={20123--20144}
}

@inproceedings{liu2019fgcnn,
  title={Feature Generation by Convolutional Neural Network for Click-Through Rate Prediction},
  author={Liu, Bin and Tang, Ruiming and Chen, Yingzhi and Yu, Jinkai and Guo, Huifeng and Zhang, Yuzhou},
  booktitle={The World Wide Web Conference},
  pages={1119--1129},
  year={2019},
  publisher={Association for Computing Machinery},
  address={New York, NY, USA}
}

@inproceedings{covington2016deep,
  title={Deep Neural Networks for YouTube Recommendations},
  author={Covington, Paul and Adams, Jay and Sargin, Emre},
  booktitle={Proceedings of the 10th ACM Conference on Recommender Systems},
  pages={191--198},
  year={2016},
  publisher={Association for Computing Machinery},
  address={New York, NY, USA}
}

@inproceedings{zhou2019dien,
  title={Deep Interest Evolution Network for Click-Through Rate Prediction},
  author={Zhou, Guorui and Mou, Na and Fan, Ying and Pi, Qi and Bian, Weijie and Zhou, Chang and Zhu, Xiaoqiang and Gai, Kun},
  booktitle={Proceedings of the AAAI Conference on Artificial Intelligence},
  pages={5941--5948},
  year={2019},
  volume={33},
  publisher={Association for the Advancement of Artificial Intelligence (AAAI)}
}

@inproceedings{kang2018sasrec,
  title={Self-Attentive Sequential Recommendation},
  author={Kang, Wang-Cheng and McAuley, Julian},
  booktitle={2018 IEEE International Conference on Data Mining (ICDM)},
  pages={197--206},
  year={2018},
  publisher={IEEE}
}

@inproceedings{sun2019bert4rec,
  title={BERT4Rec: Sequential Recommendation with Bidirectional Encoder Representations from Transformer},
  author={Sun, Fei and Liu, Jun and Wu, Jian and Pei, Changhua and Lin, Xiao and Ou, Wenwu and Jiang, Peng},
  booktitle={Proceedings of the 28th ACM International Conference on Information and Knowledge Management},
  pages={1441--1450},
  year={2019},
  publisher={Association for Computing Machinery},
  address={New York, NY, USA}
}

@inproceedings{hidasi2016gru4rec,
  title={Session-based Recommendations with Recurrent Neural Networks},
  author={Hidasi, Bal{\'a}zs and Karatzoglou, Alexandros and Baltrunas, Linas and Tikk, Domonkos},
  booktitle={International Conference on Learning Representations (ICLR)},
  year={2016}
}

@inproceedings{rendle2010fm,
  title={Factorization Machines},
  author={Rendle, Steffen},
  booktitle={2010 IEEE International Conference on Data Mining},
  pages={995--1000},
  year={2010},
  publisher={IEEE}
}

@inproceedings{cheng2016wide,
  title={Wide \& Deep Learning for Recommender Systems},
  author={Cheng, Heng-Tze and Ko{\c{c}}, Levent and Harmsen, Jeremiah and Shaked, Tal and Chandra, Tushar and Aradhye, Hrishi and Anderson, Glen and Corrado, Greg and Chai, Wei and Ispir, Mustafa and Anil, Rohan and Haque, Zakaria and Hong, Lichan and Jain, Vihan and Liu, Xiaobing and Shah, Hemal},
  booktitle={Proceedings of the 1st Workshop on Deep Learning for Recommender Systems},
  pages={7--10},
  year={2016},
  publisher={Association for Computing Machinery},
  address={New York, NY, USA}
}

@inproceedings{guo2017deepfm,
  title={DeepFM: A Factorization-Machine based Neural Network for CTR Prediction},
  author={Guo, Huifeng and Tang, Ruiming and Ye, Yunming and Li, Zhenguo and He, Xiuqiang},
  booktitle={Proceedings of the 26th International Joint Conference on Artificial Intelligence},
  pages={1725--1731},
  year={2017},
  publisher={International Joint Conferences on Artificial Intelligence Organization}
}

@inproceedings{huang2013dssm,
  title={Learning Deep Structured Semantic Models for Web Search using Clickthrough Data},
  author={Huang, Po-Sen and He, Xiaodong and Gao, Jianfeng and Deng, Li and Acero, Alex and Heck, Larry},
  booktitle={Proceedings of the 22nd ACM International Conference on Information and Knowledge Management},
  pages={2333--2338},
  year={2013},
  publisher={Association for Computing Machinery},
  address={New York, NY, USA}
}

@inproceedings{yi2019sampling,
  title={Sampling-Bias-Corrected Neural Modeling for Large Corpus Item Recommendations},
  author={Yi, Xinyang and Yang, Ji and Hong, Lichan and Cheng, Derek Zhiyuan and Heldt, Lukasz and Kumthekar, Aditee and Zhao, Zhe and Wei, Li and Chi, Ed},
  booktitle={Proceedings of the 13th ACM Conference on Recommender Systems},
  pages={269--277},
  year={2019},
  publisher={Association for Computing Machinery},
  address={New York, NY, USA}
}

@inproceedings{vaswani2017attention,
  title={Attention is All You Need},
  author={Vaswani, Ashish and Shazeer, Noam and Parmar, Niki and Uszkoreit, Jakob and Jones, Llion and Gomez, Aidan N and Kaiser, {\L}ukasz and Polosukhin, Illia},
  booktitle={Advances in Neural Information Processing Systems},
  year={2017},
  volume={30},
  publisher={Curran Associates, Inc.}
}

@inproceedings{feng2019dsin,
  title={Deep Session Interest Network for Click-Through Rate Prediction},
  author={Feng, Yufei and Lv, Fuyu and Shen, Weichen and Wang, Menghan and Sun, Fei and Zhu, Yu and Yang, Keping},
  booktitle={Proceedings of the 28th International Joint Conference on Artificial Intelligence (IJCAI)},
  pages={2301--2307},
  year={2019},
  publisher={International Joint Conferences on Artificial Intelligence Organization}
}

@inproceedings{pi2019mimn,
  title={Practice on Long Sequential User Behavior Modeling for Click-Through Rate Prediction},
  author={Pi, Qi and Bian, Weijie and Zhou, Guorui and Zhu, Xiaoqiang and Gai, Kun},
  booktitle={Proceedings of the 25th ACM SIGKDD International Conference on Knowledge Discovery \& Data Mining},
  pages={2671--2679},
  year={2019},
  publisher={Association for Computing Machinery},
  address={New York, NY, USA}
}

@inproceedings{bian2022can,
  title={CAN: Feature Co-Action Network for Click-Through Rate Prediction},
  author={Bian, Weijie and Wu, Kailun and Ren, Lejian and Pi, Qi and Zhang, Yujing and Xiao, Can and Sheng, Xiang-Rong and Zhu, Yong-Nan and Chan, Zhangming and Mou, Na and Luo, Xinchen and Xiang, Shiming and Zhou, Guorui and Zhu, Xiaoqiang and Deng, Hongbo},
  booktitle={Proceedings of the 15th ACM International Conference on Web Search and Data Mining (WSDM)},
  pages={57--65},
  year={2022},
  publisher={Association for Computing Machinery},
  address={New York, NY, USA}
}

@inproceedings{pi2020sim,
  title={Search-based User Interest Modeling with Lifelong Sequential Behavior Data for Click-Through Rate Prediction},
  author={Pi, Qi and Zhou, Guorui and Zhang, Yujing and Wang, Zhe and Ren, Lejian and Fan, Ying and Zhu, Xiaoqiang and Gai, Kun},
  booktitle={Proceedings of the 29th ACM International Conference on Information and Knowledge Management},
  pages={2685--2692},
  year={2020},
  publisher={Association for Computing Machinery},
  address={New York, NY, USA}
}

@inproceedings{krichene2020sampled,
  title={On Sampled Metrics for Item Recommendation},
  author={Krichene, Walid and Rendle, Steffen},
  booktitle={Proceedings of the 26th ACM SIGKDD International Conference on Knowledge Discovery \& Data Mining},
  pages={1748--1757},
  year={2020},
  publisher={Association for Computing Machinery},
  address={New York, NY, USA}
}

@inproceedings{chang2023twin,
  title={TWIN: TWo-stage Interest Network for Lifelong User Behavior Modeling in CTR Prediction at Kuaishou},
  author={Chang, Jianxin and Zhang, Chenbin and Fu, Zhiyi and Zang, Xiaoxue and Guan, Lin and Lu, Jing and Hui, Yiqun and Leng, Dewei and Niu, Yanan and Song, Yang and Gai, Kun},
  booktitle={Proceedings of the 29th ACM SIGKDD Conference on Knowledge Discovery and Data Mining},
  pages={3785--3794},
  year={2023},
  publisher={Association for Computing Machinery},
  address={New York, NY, USA}
}

@inproceedings{zhai2024hstu,
  title={Actions Speak Louder than Words: Trillion-Parameter Sequential Transducers for Generative Recommendations},
  author={Zhai, Jiaqi and Liao, Lucy and Liu, Xing and Wang, Yueming and Li, Rui and Cao, Xuan and Gao, Leon and Gong, Zhaojie and Gu, Fangda and He, Jiayuan and Lu, Yinghai and Shi, Yu},
  booktitle={Proceedings of the 41st International Conference on Machine Learning},
  pages={58484--58509},
  year={2024},
  volume={235},
  series={Proceedings of Machine Learning Research},
  month={21--27 Jul},
  publisher={PMLR}
}

@inproceedings{zheng2022autoattention,
  title={AutoAttention: Automatic Field Pair Selection for Attention in User Behavior Modeling},
  author={Zheng, Zuowu and Gao, Xiaofeng and Pan, Junwei and Luo, Qi and Chen, Guihai and Liu, Dapeng and Jiang, Jie},
  booktitle={2022 IEEE International Conference on Data Mining (ICDM)},
  pages={803--812},
  year={2022},
  publisher={IEEE}
}

@inproceedings{rendle2009bpr,
  title={{BPR}: {B}ayesian Personalized Ranking from Implicit Feedback},
  author={Rendle, Steffen and Freudenthaler, Christoph and Gantner, Zeno and Schmidt-Thieme, Lars},
  booktitle={Proceedings of the Twenty-Fifth Conference on Uncertainty in Artificial Intelligence (UAI)},
  pages={452--461},
  year={2009}
}

@inproceedings{kingma2014adam,
  title={{Adam}: A Method for Stochastic Optimization},
  author={Kingma, Diederik P. and Ba, Jimmy},
  booktitle={International Conference on Learning Representations (ICLR)},
  year={2015}
}

@misc{grouplens2019movielens25m,
  title={MovieLens 25M Dataset},
  author={{GroupLens Research}},
  howpublished={\url{https://grouplens.org/datasets/movielens/25m/}},
  note={Released December 2019},
  year={2019}
}

@article{ba2016layer,
  title={Layer Normalization},
  author={Ba, Jimmy Lei and Kiros, Jamie Ryan and Hinton, Geoffrey E.},
  journal={arXiv preprint arXiv:1607.06450},
  year={2016}
}

@inproceedings{reimers2019sentencebert,
  title={Sentence-{BERT}: Sentence Embeddings using Siamese {BERT}-Networks},
  author={Reimers, Nils and Gurevych, Iryna},
  booktitle={Proceedings of the 2019 Conference on Empirical Methods in Natural Language Processing and the 9th International Joint Conference on Natural Language Processing (EMNLP-IJCNLP)},
  pages={3982--3992},
  year={2019}
}

@inproceedings{wang2020minilm,
  title={MiniLM: Deep Self-Attention Distillation for Task-Agnostic Compression of Pre-Trained Transformers},
  author={Wang, Wenhui and Wei, Furu and Dong, Li and Bao, Hangbo and Yang, Nan and Zhou, Ming},
  booktitle={Advances in Neural Information Processing Systems},
  volume={33},
  pages={5776--5788},
  year={2020},
  publisher={Curran Associates, Inc.}
}

@misc{sentenceTransformers2022allMiniLML6v2,
  author={{Sentence Transformers}},
  title={{all-MiniLM-L6-v2}},
  year={2022},
  howpublished={Hugging Face model card},
  url={https://huggingface.co/sentence-transformers/all-MiniLM-L6-v2},
  note={Revision 1110a243fdf4706b3f48f1d95db1a4f5529b4d41; accessed 22 August 2026}
}

@article{srivastava2014dropout,
  title={Dropout: A Simple Way to Prevent Neural Networks from Overfitting},
  author={Srivastava, Nitish and Hinton, Geoffrey and Krizhevsky, Alex and Sutskever, Ilya and Salakhutdinov, Ruslan},
  journal={Journal of Machine Learning Research},
  volume={15},
  number={56},
  pages={1929--1958},
  year={2014}
}

@inproceedings{paszke2019pytorch,
  title={PyTorch: An Imperative Style, High-Performance Deep Learning Library},
  author={Paszke, Adam and Gross, Sam and Massa, Francisco and Lerer, Adam and Bradbury, James and Chanan, Gregory and Killeen, Trevor and Lin, Zeming and Gimelshein, Natalia and Antiga, Luca and Desmaison, Alban and K{\"o}pf, Andreas and Yang, Edward and DeVito, Zachary and Raison, Martin and Tejani, Alykhan and Chilamkurthy, Sasank and Steiner, Benoit and Fang, Lu and Bai, Junjie and Chintala, Soumith},
  booktitle={Advances in Neural Information Processing Systems},
  volume={32},
  year={2019},
  publisher={Curran Associates, Inc.}
}

@inproceedings{schnabel2016recommendations,
  title={Recommendations as Treatments: Debiasing Learning and Evaluation},
  author={Schnabel, Tobias and Swaminathan, Adith and Singh, Ashudeep and Chandak, Navin and Joachims, Thorsten},
  booktitle={Proceedings of The 33rd International Conference on Machine Learning},
  pages={1670--1679},
  year={2016},
  volume={48},
  series={Proceedings of Machine Learning Research},
  address={New York, New York, USA},
  month={20--22 Jun},
  publisher={PMLR}
}

@inproceedings{nair2010relu,
  title={Rectified Linear Units Improve Restricted Boltzmann Machines},
  author={Nair, Vinod and Hinton, Geoffrey E.},
  booktitle={Proceedings of the 27th International Conference on Machine Learning},
  pages={807--814},
  year={2010}
}

\clearpage
\appendix
\section*{Supplementary Material}

\section{Serving-Time Inference Efficiency}
\label{app:efficiency}

The teal wavy arrows in Fig.~\ref{fig:similarity-detail} mark the reusable history path: candidates in one request differ only in the target.

\noindent\textbf{Candidate-flattened evaluation and request-local reuse.} In \M{}'s cross-attention architecture, the history-encoding path is fully target-independent: the embedding lookup, projection, positional encoding, LN, and key/value projections ($\mathbf{K}, \mathbf{V}$ in Fig.~\ref{fig:similarity-detail}) depend on the history sequence only. Candidate-flattened evaluation duplicates and encodes that history for every candidate row. Request-local reuse encodes it once and expands the resulting tensors across the $C$ candidates. Both paths discard their tensors after the call. The remaining per-candidate computation consists of the attention scores $\boldsymbol{\alpha}_h$, pooling $\mathbf{u}_h$, element-wise cross $\mathbf{c}_h = \mathbf{u}_h \odot \boldsymbol{\eta}_h$, identity branch, and scorer. A target-appending Transformer has no such complete target-independent path.

\noindent\textbf{Pre-computation boundary.} Let $\mathcal{I}$ denote the item universe. Frozen embeddings make the target-side $\mathbf{Q}$ and target-map inputs independent of the current history. History $\mathbf{K}$ and $\mathbf{V}$ are different: because positional encoding is added before LN and the key/value projections, they depend on both item and sequence position. Exact request-local reuse therefore computes them after assembling the positioned history. Storing item--position representations would require $O(|\mathcal{I}|SD_u)$ rather than $O(|\mathcal{I}|D_u)$ space. The aggressive industrial benchmark in Supp.~Sec.~\ref{app:latency-detail} excludes this assembled-history computation and should be read as a lower bound on online latency, not as an item-only lookup.

\section{Analytical FLOP Accounting}
\label{app:flops}

Under history-linearity, PS reduces calls to $\phi$ from $S$ to one per candidate.
Cross-attention separately makes the key and value paths target-independent and reusable
across candidates. We count FLOPs for scoring $C = 100$ candidates against a history of
$S$ items, including arithmetic in the scoring path (projections, attention, cross
networks, MLP) but excluding embedding-table lookups. The candidate-flattened path
repeats history encoding for each candidate. Request-local reuse evaluates it once and includes
that one evaluation in the total. Neither path retains state across requests.

For the controlled calculation at $C=100$, we isolate the reusable history path. The measured comparison across candidate counts appears in Tab.~\ref{tab:latency-public}.

\begin{table*}[t]
\centering
\small
\caption{Total FLOPs for scoring $C = 100$ candidates in the shared public setting ($S = 50$, $D = 64$). ``Flat.'' re-encodes the shared history for every candidate, while ``Reuse'' encodes it once within the request. The benefit is Flat./Reuse.}
\label{tab:flops-public}
\setlength{\tabcolsep}{4pt}
\begin{tabular}{@{}lrrr@{}}
\toprule
Model & Flat. & Reuse & Benefit \\
\midrule
DIN & $421.38$M & $176.17$M & $2.39\times$ \\
DIN + DCN-v2 & $429.04$M & $183.84$M & $2.33\times$ \\
\M{} (1 layer, 4 experts) & $618.50$M & $43.83$M & $14.11\times$ \\
\M{} (2 layers, 4 experts) & $635.56$M & $60.88$M & $10.44\times$ \\
\M{} (1 layer, 8 experts) & $635.42$M & $60.74$M & $10.46\times$ \\
\bottomrule
\end{tabular}
\end{table*}

After request-local reuse, the prespecified \M{} configuration uses $43.83$M FLOPs,
one quarter of DIN's $176.17$M. Re-encoding the history for each candidate raises
\M{} to $618.50$M FLOPs, so sharing that path yields a $14.11\times$ analytical
benefit at $C=100$. DIN obtains a smaller $2.39\times$ benefit because its
target-independent embedding, projection, and LN can be shared, while its activation
unit remains jointly target--history dependent. A target-augmented Transformer has an
analytical benefit of $1.00\times$ because its forward pass expands the candidates
before self-attention.

At $C=1{,}000$, the analytical \M{} benefit is $16.02\times$, compared with measured
medians of $14.21\times$ on Steam and $14.68\times$ on ML-25M. At smaller candidate
counts, fixed call overhead occupies a larger share of the reused path, so the measured
wall-clock benefit is below the FLOP ratio.

\noindent\textbf{Counting convention.} We count \emph{multiply--accumulate}~(MAC) operations and element-wise operations (softmax, LN). We then apply the $\text{FLOPs} = 2 \times \text{MACs}$ convention to their sum, which double-counts the non-MAC terms. These terms are a small fraction treated identically across models, so ratios are less affected than absolute totals, which are estimates, not instruction counts.

\section{\M{} Architecture Equations and Cross-Layer Comparison}
\label{app:cross-comparison}

This supplementary section gives full equations for all five stages of \M{}'s forward pass (cross-referenced from Sec.~\ref{sec:architecture}) and compares the cross layer in Stage~3 against a standard DCN-v2 self-cross. The main-text notation is summarised in Tab.~\ref{tab:notation}. Additional batched quantities are defined below when introduced. Fig.~\ref{fig:similarity-detail} gives a detailed view of the similarity branch data flow.

\subsection*{Stage 1: Embedding and Normalisation}

Given raw frozen embeddings $\mathbf{x}_i^{\text{raw}}, \mathbf{t}_c^{\text{raw}}$ of an arbitrary source dimension, the shared projection and LN produce $\tilde{\mathbf{X}}\in\mathbb{R}^{S\times D}$ and $\tilde{\mathbf{T}}\in\mathbb{R}^{C\times D}$, with rows
\begin{align}
    \tilde{\mathbf{x}}_i
        &= \mathrm{Dropout}\bigl(\mathrm{LN}(\mathrm{proj}(\mathbf{x}_i^{\text{raw}}) + \mathrm{pos}_i)\bigr) \in \mathbb{R}^D, \\
    \tilde{\mathbf{t}}_c
        &= \mathrm{LN}(\mathrm{proj}(\mathbf{t}_c^{\text{raw}})) \in \mathbb{R}^D.
\end{align}

\subsection*{Stage 2: Multi-Head Dot-Product Cross-Attention}

The projections map width $D$ to $D_u=Hd$, which need not equal $D$:
\begin{gather}
    \mathbf{Q} = \mathrm{reshape}_H(\tilde{\mathbf{T}}\mathbf{W}_q^\top) \in \mathbb{R}^{H \times C \times d}, \\
    \mathbf{K} = \mathrm{reshape}_H(\tilde{\mathbf{X}}\mathbf{W}_k^\top) \in \mathbb{R}^{H \times S \times d}, \\
    \mathbf{V} = \mathrm{reshape}_H(\tilde{\mathbf{X}}\mathbf{W}_v^\top) \in \mathbb{R}^{H \times S \times d}, \\
    \mathbf{V}_t = \mathrm{reshape}_H(\tilde{\mathbf{T}}\mathbf{W}_v^\top) \in \mathbb{R}^{H \times C \times d}, \\
    \boldsymbol{\alpha}_h
        = \mathrm{softmax}\!\left(\frac{\mathbf{Q}_h \mathbf{K}_h^\top}{\sqrt{d}}\right), \\
    \mathbf{u}_h
        = \boldsymbol{\alpha}_h \mathbf{V}_h.
\end{gather}
Here $\boldsymbol{\alpha}_h\in\mathbb{R}^{C\times S}$ and $\mathbf{u}_h\in\mathbb{R}^{C\times d}$. The row of $\mathbf{V}_t$ for one candidate is denoted $\mathbf{v}_t=[\mathbf{v}_t^{(1)};\ldots;\mathbf{v}_t^{(H)}]\in\mathbb{R}^{D_u}$. Stage 3 operates independently on these head slices.

\subsection*{Stage 3a: Comparison with DCN-v2}

The Stage 3 similarity operator is structurally related to DCN-v2 but differs along several axes. Tab.~\ref{tab:cross-comparison} summarises the differences.

\begin{table*}[t]
\centering
\small
\caption{Structural comparison of our user--target cross layer (in \M{}'s similarity branch) against a standard DCN-v2 self-cross.}
\label{tab:cross-comparison}
\begin{tabular}{@{}p{2.7cm}p{6.5cm}p{6.5cm}@{}}
\toprule
\textbf{Aspect} & \textbf{DCN-v2 self-cross} & \textbf{Our user--target cross} \\
\midrule
Inputs & One feature vector $\mathbf{x}$, self-crossed. & Pooled history $\mathbf{u}\in\mathbb{R}^{D_u}$ and target $\tilde{\mathbf{t}}\in\mathbb{R}^D$, with a target projection $\mathbf{v}_t\in\mathbb{R}^{D_u}$. \\[4pt]
Core operation & $\mathbf{x}_{\ell+1} = \mathbf{x}_0 \odot (\mathbf{W}_\ell \mathbf{x}_\ell + \mathbf{b}_\ell) + \mathbf{x}_\ell$ & $\mathbf{c} = \mathbf{u} \odot \boldsymbol{\eta}(\tilde{\mathbf{t}})$ (no residual), with $\boldsymbol{\eta}$ computed from $\mathbf{v}_t$ and gates on $\tilde{\mathbf{t}}$. \\[4pt]
Residual connection & Included, carries the previous layer state $\mathbf{x}_\ell$ forward. & Dropped. The output contains only similarity, and the identity branch handles target-only content. \\[4pt]
Head structure & Single-stream, no attention heads. & Each head operates on $d$-dimensional slices of $\mathbf{u}$ and $\mathbf{v}_t$; concatenation gives $D_u=Hd$. \\[4pt]
Expert structure & Mixture of experts is global: $\mathbf{M}_\ell(\mathbf{x}_\ell) = \sum_k g_{\ell,k}(\mathbf{x}_\ell)\mathbf{U}_{\ell,k}\mathbf{V}_{\ell,k}$, dense across all dimensions. & The same $d$-dimensional experts are reused across heads, producing $H$ independent blocks $\mathbf{M}_h(\tilde{\mathbf{t}})$. Relative to a width-$D$ DCN-v2 layer, the expert-term ratio is $d/D$. \\[4pt]
Gating & Single gate vector $\mathbf{g}_\ell(\mathbf{x}_\ell) \in \mathbb{R}^K$ shared across all dimensions. & Per-head gate vector $\mathbf{g}_h(\mathbf{t}) \in \mathbb{R}^K$, so each head mixes the $K$ experts independently instead of sharing one mixture across all dimensions. \\[4pt]
Depth & Cross layers can be stacked to model higher-order feature interactions. & The history-linearity and PS-equivalence analysis applies to the single-layer operator (Sec.~\ref{sec:examples}). Stacked variants fall outside this guarantee and change the model class. \\[4pt]
Role in the pipeline & Feature-interaction module on flat features. & Instantiation of the similarity operator $\phi$ of the similarity--identity decoupling framework (Sec.~\ref{sec:framework}). \\
\bottomrule
\end{tabular}
\end{table*}

The two operators share an element-wise product and a low-rank expert mixture. Our two-input form, residual removal, multi-head block-diagonal structure, per-head gating, and single-layer restriction together define our history-linear similarity operator.

\subsection*{Stage 3b: Gated-Cross Equations}

We now write the cross layer used in \M{}'s similarity branch alongside a single-layer DCN-v2 self-cross. The Stage 1 target is $\tilde{\mathbf{t}}\in\mathbb{R}^D$. Attention uses $H$ heads of width $d$, so the pooled history is $\mathbf{u}=[\mathbf{u}_1;\ldots;\mathbf{u}_H]\in\mathbb{R}^{D_u}$ with $D_u=Hd$. Stage 3 uses the target value projection $\mathbf{v}_t=\mathbf{W}_v\tilde{\mathbf{t}}\in\mathbb{R}^{D_u}$, partitioned into $\mathbf{v}_t^{(h)}\in\mathbb{R}^d$. The gate for each head still receives the full $\tilde{\mathbf{t}}$.

\noindent\textbf{DCN-v2 self-cross on the target (reference operator).} A single DCN-v2 cross layer applied to the target alone computes
\begin{align}
    \mathbf{g}(\tilde{\mathbf{t}}) &= \mathrm{softmax}(\mathbf{W}_{\text{gate}} \tilde{\mathbf{t}}) \in \mathbb{R}^K, \\
    \mathbf{M}(\tilde{\mathbf{t}}) &= \sum_{k=1}^K g_k(\tilde{\mathbf{t}}) \, \mathbf{U}_k \mathbf{V}_k \in \mathbb{R}^{D \times D}, \label{eq:dcn-m} \\
    \tilde{\mathbf{t}}_{\text{crossed}} &= \tilde{\mathbf{t}} \odot \bigl(\mathbf{M}(\tilde{\mathbf{t}}) \, \tilde{\mathbf{t}} + \mathbf{b}\bigr) + \tilde{\mathbf{t}} \in \mathbb{R}^D,
\end{align}
where $\mathbf{U}_k \in \mathbb{R}^{D \times r}$ and $\mathbf{V}_k \in \mathbb{R}^{r \times D}$ are global (non-head-structured) low-rank experts. The gating $\mathbf{g}$ is a single shared vector, and the residual connection $+\tilde{\mathbf{t}}$ re-injects the projected target.

\noindent\textbf{\M{}'s user--target cross layer.} We apply the analogous operator to the two distinct inputs $(\mathbf{u}, \tilde{\mathbf{t}})$, per head, without a residual, and with per-head gating over block-diagonal experts:
\begin{align}
    \mathbf{g}_h(\tilde{\mathbf{t}}) &= \mathrm{softmax}(\mathbf{W}_{g,h} \, \tilde{\mathbf{t}}) \in \mathbb{R}^K, & h = 1, \ldots, H, \\
    \mathbf{f}_{h,k} &= \mathbf{U}_k \mathbf{V}_k \, \mathbf{v}_t^{(h)} \in \mathbb{R}^d, & k = 1, \ldots, K, \label{eq:melodi-expert} \\
    \mathbf{c}_h &= \mathbf{u}_h \odot \Bigl( \sum_{k=1}^K g_{h,k}(\tilde{\mathbf{t}}) \, \mathbf{f}_{h,k} + \mathbf{b}_h \Bigr) \in \mathbb{R}^d, \label{eq:melodi-ch} \\
    \mathbf{c} &= \mathrm{concat}(\mathbf{c}_1, \ldots, \mathbf{c}_H) \in \mathbb{R}^{D_u}, \label{eq:melodi-c}
\end{align}
where $\mathbf{U}_k \in \mathbb{R}^{d \times r}$ and $\mathbf{V}_k \in \mathbb{R}^{r \times d}$ are rectangular low-rank experts shared across heads and applied independently within each $d$-dimensional head subspace.

\noindent\textbf{Compact form.} Equations~\eqref{eq:melodi-expert}--\eqref{eq:melodi-ch} can equivalently be written as $\mathbf{c} = \mathbf{u} \odot \boldsymbol{\eta}(\tilde{\mathbf{t}})$ with the block-diagonal target map
\begin{align}
    \boldsymbol{\eta}(\tilde{\mathbf{t}}) &= \bigl[ \mathbf{M}_1(\tilde{\mathbf{t}}) \, \mathbf{v}_t^{(1)}+\mathbf{b}_1; \, \ldots; \, \mathbf{M}_H(\tilde{\mathbf{t}}) \, \mathbf{v}_t^{(H)}+\mathbf{b}_H \bigr] \in \mathbb{R}^{D_u}, \label{eq:melodi-eta} \\
    \mathbf{M}_h(\tilde{\mathbf{t}}) &= \sum_{k=1}^K g_{h,k}(\tilde{\mathbf{t}}) \, \mathbf{U}_k \mathbf{V}_k \in \mathbb{R}^{d \times d}. \label{eq:melodi-mh}
\end{align}

\noindent\textbf{Vectorised evaluation.} The per-head operations can be evaluated as batched tensor contractions. Let $N=BC$ combine the $B$ examples and $C$ candidates, and let $\mathbf{V}_{t,H} \in \mathbb{R}^{N \times H \times d}$ denote the projected targets partitioned into heads. Within this paragraph, $\mathbf{u}$, $\boldsymbol{\eta}$, and $\mathbf{c}$ denote stacked tensors in $\mathbb{R}^{N\times D_u}$. For expert $k$, rank coordinate $\ell$, head $h$, batch--candidate index $n$, and head coordinates $j,q$, define
\begin{align}
    \mathcal{Z}_{k\ell hn}
        &= \sum_{q=1}^{d} (\mathbf{V}_k)_{\ell q}(\mathbf{V}_{t,H})_{nhq},
        & \mathcal{Z} &\in \mathbb{R}^{K \times r \times H \times N}, \label{eq:vec-vx} \\
    \mathcal{E}_{kjhn}
        &= \sum_{\ell=1}^{r}(\mathbf{U}_k)_{j\ell}\mathcal{Z}_{k\ell hn},
        & \mathcal{E} &\in \mathbb{R}^{K \times d \times H \times N}. \label{eq:vec-experts}
\end{align}
The $K$ expert matrices are stacked as $\mathbf{V}_{1:K} \in \mathbb{R}^{K \times r \times d}$ and $\mathbf{U}_{1:K} \in \mathbb{R}^{K \times d \times r}$. After re-indexing $\mathcal{E}$ as $\mathbf{E}\in\mathbb{R}^{N \times K \times D_u}$, the per-head gates~$\mathbf{G} \in \mathbb{R}^{N \times K \times D_u}$, with each $g_{h,k}$ repeated across its $d$ coordinates, weight and sum the expert outputs:
\begin{equation}
    \boldsymbol{\eta} = \textstyle\sum_{k=1}^{K} \mathbf{E}_{:,k,:} \odot \mathbf{G}_{:,k,:} \;+\; \mathbf{b} \;\in\; \mathbb{R}^{N \times D_u}, \quad
    \mathbf{c} = \mathbf{u} \odot \boldsymbol{\eta} \;\in\; \mathbb{R}^{N \times D_u}. \label{eq:vec-cross}
\end{equation}

The per-head target matrix $\mathbf{M}_h(\tilde{\mathbf{t}})$ is a softmax-gated mixture of rank-$r$ experts. Their block-diagonal composition has shape $\mathrm{blkdiag}(\mathbf{M}_1,\ldots,\mathbf{M}_H)\in\mathbb{R}^{D_u\times D_u}$, whereas DCN-v2 uses a dense width-$D$ matrix with global gating. The expert weight matrices contain $2Kdr$ and $2KDr$ weights, respectively, giving a ratio of $d/D$. The ratio equals $1/H$ only when $D_u=D$.

\noindent\textbf{Weight count.} Counting weight matrices and excluding biases, a width-$D$ DCN-v2 self-cross uses $K(2Dr+D)$ weights for experts and gates. Our operator uses $K(2dr+HD)$ because the experts are shared across heads and each of the $H$ gates sees the full target. It is smaller exactly when $2r(D-d)>D(H-1)$. This condition need not hold because $d$ is configurable. If $D_u=D$, then $d=D/H$ and the condition reduces to $2r>H$. For $H{=}4$, $d{=}64$, $r{=}32$, $K{=}4$, and $D{=}256$, the weight counts are $20{,}480$ and $66{,}560$. In the public experiments, by contrast, $d=D$ and $D_u=4D$, so the block structure does not reduce the expert term relative to a width-$D$ DCN-v2 layer and the per-head gates add weights.

\subsection*{Stage 4: Identity Branch}

The identity branch is problem-dependent (Sec.~\ref{sec:identity-branch}). In our experiments, we use a DCN-v2 self-cross on the target, whose equations are given in the ``DCN-v2 self-cross on the target'' paragraph above. Thus $\tilde{\mathbf{t}}_{\mathrm{id}} \triangleq \tilde{\mathbf{t}}_{\text{crossed}} \in \mathbb{R}^D$.

\subsection*{Stage 5: MLP Scoring}

In the experiments we adopt the cross-only form, in which the MLP consumes the similarity-branch output $\mathbf{c}$ and the identity-branch output $\tilde{\mathbf{t}}_{\mathrm{id}}$:
\begin{equation}
    \hat{y} = \mathrm{MLP}\bigl([\mathbf{c}\,;\,\tilde{\mathbf{t}}_{\mathrm{id}}]\bigr) \in \mathbb{R}.
\end{equation}
The scorer uses a hidden layer of width $D$ with a \emph{rectified linear unit} activation~\citep{nair2010relu}, followed by a scalar output layer.
Ablation~\#5 in Supp.~Sec.~\ref{sec:ablations} adds the uncrossed pooled history $\mathbf{u}$ and projected target $\tilde{\mathbf{t}}$ to this input. MRR changes by $+0.0038$ on Steam and $+0.0005$ on ML-25M relative to the cross-only scorer.

\section{Ablations on the Public Datasets}
\label{sec:ablations}

Tab.~\ref{tab:ablations} reports ablations of the 4-expert configuration. The anchor is the single-layer, residual-free configuration whose MLP receives only the crossed outputs, with the identity branch enabled. Rows~\#3b and \#4b isolate the pooled-history residual and similarity-depth interventions, and rows~\#3a and \#4a are the coupled variants.

\begin{itemize}[leftmargin=*,nosep]
\item \textbf{Observed ranges.} Across the tested interventions, Steam changes range from $-0.0077$ to $+0.0074$, while ML-25M changes range from $-0.0151$ to $+0.0053$.
\item \textbf{\#1 vs.\ \#2 (identity branch).} Removing the target self-cross changes MRR by $-0.0018$ on Steam and $-0.0151$ on ML-25M. The larger observed ML-25M drop motivates retaining the tested identity branch (Sec.~\ref{sec:identity-branch}).
\item \textbf{\#1 vs.\ \#3 (residual connection).} Two forms are reported because the evaluated recurrence adds its \emph{expert input} as the residual. In the first layer, this input is the target rather than the pooled history $\mathbf{u}$, and no residual is defined when the expert-input width differs from $D_u$. Row~\#3b instead adds an explicit $+\mathbf{u}$ term. Let $\mathbf{1}\in\mathbb{R}^{D_u}$ denote the all-ones vector. This term remains history-linear because $\mathbf{u}\odot\boldsymbol{\eta}(\mathbf{t})+\mathbf{u}=\mathbf{u}\odot(\boldsymbol{\eta}(\mathbf{t})+\mathbf{1})$. Target and pooled-history residuals change MRR by $+0.0002$ and $-0.0004$ on Steam. Both changes are $-0.0010$ on ML-25M.
\item \textbf{\#1 vs.\ \#4 (cross depth).} Row~\#4b isolates similarity depth by holding the identity branch at one layer and four experts. A second similarity layer changes MRR by $+0.0074$ on Steam and $-0.0014$ on ML-25M, so the accuracy direction is dataset-dependent. Row~\#4a also deepens the identity branch and changes MRR by $+0.0051$ and $+0.0053$, respectively. The SP/PS guarantee of Sec.~\ref{sec:examples} applies to the single-layer operator, while the stacked variant changes the model class. We therefore retain one layer.
\item \textbf{\#1 vs.\ \#5 (uncrossed inputs in MLP).} Adding the uncrossed pooled history $\mathbf{u}$ and projected target $\tilde{\mathbf{t}}$ to the crossed features changes MRR by $+0.0038$ on Steam and $+0.0005$ on ML-25M.
\item \textbf{\#1 vs.\ \#6 (projected target only).} Including only $\tilde{\mathbf{t}}$ changes MRR by $-0.0077$ on Steam and $-0.0099$ on ML-25M. Row~\#5 exceeds row~\#6 on both datasets, so the observed values favour accompanying the projected target with the uncrossed pooled history it is meant to be compared against.
\end{itemize}

\subsection*{Expert Count}

\begin{table}[h]
\centering
\small
\caption{Effect of doubling expert count. $\Delta$ is MRR with 8 experts minus MRR with 4 experts. All variants are residual-free, use cross-only scoring, and include the target self-cross. ``Pinned'' holds the identity signal branch at one layer and 4 experts, while ``Both'' changes both signal branches.}
\label{tab:experts}
\begin{tabular}{@{}lcccc@{}}
\toprule
& \multicolumn{2}{c}{Steam} & \multicolumn{2}{c}{ML-25M} \\
\cmidrule(lr){2-3} \cmidrule(lr){4-5}
Configuration & Both & Pinned & Both & Pinned \\
\midrule
1 similarity layer & $+0.0088$ & $+0.0035$ & $+0.0048$ & $-0.0032$ \\
2 similarity layers & $+0.0030$ & $-0.0088$ & $+0.0042$ & $+0.0031$ \\
\bottomrule
\end{tabular}
\end{table}

With identity capacity fixed, doubling expert count changes MRR by $+0.0035$ and $-0.0088$ on Steam and by $-0.0032$ and $+0.0031$ on ML-25M for one and two layers, respectively. The direction differs across depths and datasets. The coupled variants improve MRR in all four comparisons, showing that identity-branch changes can alter the apparent effect of expert count. We retain four experts as a fixed default that these single-run comparisons do not select.

\subsection*{Fixed Defaults}

Tab.~\ref{tab:recommendations} lists the fixed defaults, effects, and recommendations.

\begin{table*}[t]
\centering
\small
\caption{Fixed defaults and their observed MRR effects. Each effect compares single training runs.}
\label{tab:recommendations}
\begin{tabular}{@{}p{4.7cm}p{3.0cm}p{7.9cm}@{}}
\toprule
Component & Effect (Steam / ML-25M) & Recommendation \\
\midrule
Remove target self-cross from identity branch & $-0.0018$ / $-0.0151$ & Retain the tested identity branch because removal gives lower reported MRR \\
Pooled-history residual & $-0.0004$ / $-0.0010$ & Omit to keep a pure interaction signal \\
Second similarity layer & $+0.0074$ / $-0.0014$ & Use one layer to retain the stated SP/PS guarantee \\
Uncrossed pooled history and projected target in MLP & $+0.0038$ / $+0.0005$ & Retain the fixed cross-only scorer \\
Projected target without uncrossed pooled history & $-0.0077$ / $-0.0099$ & If the scorer receives uncrossed inputs, pair the target with pooled history \\
Expert count & Changes sign across settings & Use 4 experts as the fixed default \\
\bottomrule
\end{tabular}
\end{table*}

\section{Capacity-Matched TA-Transformer}
\label{app:capacity-ta}

The TA-Transformer in Tab.~\ref{tab:main} uses one layer and one head to match the industrial comparator. To separate architecture from capacity, we train two-layer, two-head variants with feed-forward widths $2D$ (95{,}041 trainable parameters) and $4D$ (128{,}065), the latter close to \M{}'s 132{,}885.
All other settings match the public protocol: learning rate $10^{-3}$, $D=64$, dropout $0.1$, training batch size $1{,}024$, evaluation batch size $256$, at most $200$ epochs, patience $10$, and $S=50$. We train each variant once except the parameter-matched Steam model, which uses three seeds.
Latency uses request-local reuse at $C=1{,}000$ on the same 32-core processor. We disable PyTorch's nested-tensor path, which made the two-head models about $4.5\times$ slower for this request shape; disabling it changes no score by more than $5\times10^{-8}$.

\begin{table*}[t]
\centering
\scriptsize
\caption{Larger TA-Transformers: single-run metrics with $99$ sampled negatives and request-local latency at $C=1{,}000$ (ms; Steam / ML-25M). Latency is from one benchmark run.}
\label{tab:capacity-ta}
\begin{tabular}{@{}lrccccccc@{}}
\toprule
& & \multicolumn{3}{c}{Steam} & \multicolumn{3}{c}{ML-25M} & \\
Model & Params & HIT@10 & NDCG@10 & MRR & HIT@10 & NDCG@10 & MRR & Latency \\
\midrule
TA-Transformer (1L, 1H) & 61.6k & .7695 & .5073 & .4370 & .9302 & .7189 & .6546 & 15.0 / 17.1 \\ %
TA-Transformer (2L, 2H) & 95.0k & .7960 & .5421 & .4727 & .9407 & .7399 & .6783 & 29.3 / 29.0 \\ %
TA-Transformer (2L, 2H, $4D$) & 128.1k & \textbf{.8080} & \textbf{.5562} & \textbf{.4867} & .9461 & .7519 & .6921 & 32.4 / 31.3 \\ %
\M{} & 132.9k & .7923 & .5405 & .4720 & \textbf{.9510} & \textbf{.7619} & \textbf{.7033} & \textbf{2.0 / 2.0} \\ %
\bottomrule
\end{tabular}
\end{table*}

On Steam, the parameter-matched model exceeds \M{} on all six sampled metrics in each of three seeds (MRR $0.4821\pm0.0048$ versus $0.4720\pm0.0016$). In one run, with $999$ retrieved negatives, it also leads (\M{} minus it: $-0.0040$ [$-0.0046$, $-0.0035$] MRR), but under full-catalogue ranking \M{} leads by $+0.0026$ [$+0.0022$, $+0.0031$].
On Steam, the smaller two-head model ties \M{} in sampled MRR and has lower MRR under hard and full ranking.
On ML-25M, the parameter-matched model's sampled MRR of $0.6921$ is above DIN + DCN-v2 at $0.6850$, while \M{} remains higher on every reported metric and leads it by $+0.0113$, $+0.0043$, and $+0.0083$ MRR under sampled, hard, and full ranking, with all paired-bootstrap $95\%$ intervals above zero.
At $C=1{,}000$, the parameter-matched model is about $16\times$ slower than \M{} because it reruns target-augmented self-attention for every candidate.

\section{Latency Benchmarks}
\label{app:latency-detail}

The public and industrial latency benchmarks use the same commodity processor instance with 32 virtual cores and no graphics-processor acceleration: an x86 platform with 64\,GiB of memory, backed by third-generation Intel Xeon Scalable (Ice Lake) processors. The headline measurements score one user with 50 history items against 1{,}000 candidates.

\noindent\textbf{Public datasets.} The public sweep described in~Sec.~\ref{sec:public-evaluation} also varies the candidate count. Every model runs with PyTorch 2.1~\citep{paszke2019pytorch} in its own fresh process under the same software image, thread count, and allocator settings. Five fixed permuted model orders vary each model's position across replicates. Each path records the minimum of 300 timed calls after 30 warm-up calls. Candidate-flattened and request-local-reuse paths run in the same process with separate warm-up blocks, and their order alternates across replicates. Their scores agree within a maximum absolute difference of $2.2\times10^{-8}$. Medians across replicate minima appear in Tab.~\ref{tab:latency-public}. At $C=1{,}000$, the maximum-to-minimum spread is at most $1.04\times$ for both \M{} paths and reaches $2.56\times$ for DIN's reused path. Across replicates, the paired \M{} benefit at $C=1{,}000$ spans $13.90$--$14.50\times$ on Steam and $14.47$--$14.98\times$ on ML-25M. A second run with $C\in\{1,10,50,100,150,200,300,500,700,1{,}000,2{,}000\}$ reproduces every cell at $C\leq100$ within $0.02$\,ms and \M{}'s latency at $C=1{,}000$ ($1.96$ and $1.95$\,ms). DIN's reused path is faster than \M{}'s up to $C=500$ ($1.49$ versus $1.64$\,ms on both datasets) and slower from $C=700$ ($5.08$ versus $1.71$\,ms on Steam and $2.16$ versus $1.72$\,ms on ML-25M), and DIN + DCN-v2 becomes slower than \M{} between $C=200$ and $300$. Baseline latencies at $C=1{,}000$ differ from Tab.~\ref{tab:latency-public} by 9--25\%, for example $7.38$ versus $9.89$\,ms for DIN on Steam, so ratios against baselines depend on the run.

\noindent\textbf{Industrial datasets.} The industrial benchmark uses five warm-up and 20 timed iterations and reports their mean. Its candidate-flattened path batches 1{,}000 candidates with a duplicated history and therefore re-encodes that history for every candidate. The request-local-reuse path first processes one history through embedding lookup and projection, positional encoding, LN, $\mathbf{W}_k$, and $\mathbf{W}_v$, together with padding and event masks and shared request features. It then expands those tensors while batching the target embedding and query path, attention, time decay, event-history pooling, cross modules, identity signal branch, and scorer across candidates. Thus the reported $2.6$--$3.8\times$ reduction compares candidate-flattened re-encoding with explicit single-history reuse in the same request. The fixed history length standardises the comparison but does not cover all observed industrial histories.

\begin{figure*}[t]
\centering
\includegraphics{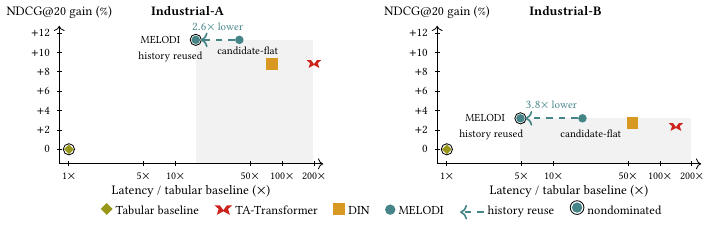}
\caption{Industrial quality gain versus latency relative to the tabular baseline on a commodity processor instance with 32 virtual cores. Quality is the relative NDCG@20 gain (\%) over the tabular baseline. The logarithmic horizontal axis is model latency divided by tabular-baseline latency, so the upper left is better. In both datasets, \M{} is faster and has higher reported NDCG@20 than TA-Transformer and DIN. Blue dashed arrows show the measured candidate-flattened-to-request-local-reuse reduction, grey regions are dominated by reused \M{}, and black rings mark nondominated measured operating points.}
\Description{Two scatter plots, one per industrial dataset, of relative NDCG@20 gain (percent) on
  the vertical axis against latency overhead on a logarithmic horizontal axis,
  both measured relative to the tabular baseline. In each panel, MELODI sits
  above and to the left of both TA-Transformer and DIN. A dashed
  arrow points left from candidate-flattened to request-local-reuse MELODI.
  Black rings identify the tabular baseline and reused MELODI as the
  nondominated measured points.}
\label{fig:pareto}
\end{figure*}

\begin{table}[h]
\centering
\small
\caption{Industrial-A latency on a commodity processor instance with 32 virtual cores. Latency is divided by tabular-baseline latency, and NDCG@20 gain is the percentage change relative to the tabular baseline.}
\label{tab:latency-a}
\begin{tabular}{@{}lrr@{}}
\toprule
Model & Latency ($\times$) & NDCG@20 gain (\%) \\
\midrule
Tabular baseline & \textbf{1.0} & -- \\
TA-Transformer & 201.8 & +8.8 \\
DIN & 80.9 & +8.8 \\
\M{} (candidate-flattened) & 40.0 & \textbf{+11.3} \\
\M{} (history reused) & 15.6 & \textbf{+11.3} \\
\bottomrule
\end{tabular}
\end{table}

\noindent\textbf{Industrial-A.} Reuse reduces \M{}'s latency from $40.0\times$ to $15.6\times$ the tabular baseline, a $2.6\times$ reduction. The reused model is $13.0\times$ faster than TA-Transformer and $5.2\times$ faster than DIN.

\begin{table}[h]
\centering
\small
\caption{Industrial-B latency on a commodity processor instance with 32 virtual cores, reported as in Tab.~\ref{tab:latency-a}.}
\label{tab:latency-b}
\begin{tabular}{@{}lrr@{}}
\toprule
Model & Latency ($\times$) & NDCG@20 gain (\%) \\
\midrule
Tabular baseline & \textbf{1.0} & -- \\
TA-Transformer & 142.1 & +2.3 \\
DIN & 55.1 & +2.7 \\
\M{} (candidate-flattened) & 18.7 & \textbf{+3.2} \\
\M{} (history reused) & 4.9 & \textbf{+3.2} \\
\bottomrule
\end{tabular}
\end{table}

\noindent\textbf{Industrial-B.} Reuse reduces \M{}'s latency from $18.7\times$ to $4.9\times$ the tabular baseline, a $3.8\times$ reduction. The reused model is $28.7\times$ faster than TA-Transformer and $11.1\times$ faster than DIN. Industrial-B uses $d{=}32$, whereas Industrial-A uses $d{=}64$. The configurations also differ in expert count and the cross between the event-history branches. TA-Transformer remains slowest in both industrial datasets because each candidate requires self-attention over a separate target--history sequence.

\subsection*{Aggressive pre-computation lower bound}

Tab.~\ref{tab:latency-offline} reports an aggressive lower bound that excludes construction of positioned history $\mathbf{K},\mathbf{V}$ and candidate $\mathbf{Q}$ and normalised target representations. The Industrial-A timed path retains masks, time decay, shared-feature projections, attention, pooling, cross modules, and the scorer. The Industrial-B timing protocol also moves the masks, time-decay weights, and shared-feature projections outside the timed region. Because positioned $\mathbf{K},\mathbf{V}$ cannot be recovered from an item-only lookup, both benchmarks assume they have already been computed for the assembled history. Their change from request-local reuse is below $10\%$ and has no consistent direction.

\begin{table}[h]
\centering
\small
\caption{Online-latency lower bound with aggressive pre-computation on a commodity processor instance with 32 virtual cores ($1$ user $\times$ $50$ history $\times$ $1{,}000$ candidates). Latencies are divided by the tabular-baseline latency of each dataset.}
\label{tab:latency-offline}
\begin{tabular}{@{}lrrr@{}}
\toprule
Dataset / Model & Flat. & Reuse & Lower bound \\
\midrule
Industrial-A: \M{} & $40.0\times$ & $15.6\times$ & $15.7\times$ \\
Industrial-B: \M{} & $18.7\times$ & $4.9\times$ & $4.6\times$ \\
\bottomrule
\end{tabular}
\end{table}

\section{Experimental Details}
\label{app:experimental-details}

\begin{table}[t]
\centering
\small
\caption{Trainable parameters, excluding the frozen item-embedding table, and configurations trained per public dataset. Counts are identical on Steam and ML-25M.}
\label{tab:params}
\setlength{\tabcolsep}{4pt}
\begin{tabular}{@{}lrr@{}}
\toprule
Model & Parameters & Configurations \\
\midrule
SASRec & 127{,}936 & 2 \\
TA-Transformer (1 layer, 1 head) & 61{,}569 & 1 \\
TA-Transformer (2 layers, 2 heads) & 95{,}041 & 1 \\ %
TA-Transformer (2 layers, 2 heads, $4D$) & 128{,}065 & 1 \\ %
DIN & 52{,}802 & 1 \\
DIN + DCN-v2 & 90{,}314 & 1 \\
CAN & 109{,}437 & 1 \\ %
DIEN & 90{,}113 & 1 \\ %
\M{} (4 experts, 1 layer) & 132{,}885 & \multirow{5}{*}{13} \\
\M{} (4 experts, 2 layers) & 167{,}273 & \\
\M{} (8 experts, 1 layer) & 166{,}953 & \\
\M{} (8 experts, 2 layers) & 235{,}409 & \\
\M{} without similarity cross & 115{,}205 & \\ %
\bottomrule
\end{tabular}
\end{table}

\begin{table*}[t]
\centering
\small
\caption{Dataset statistics and evaluation protocols.}
\label{tab:datasets}
\begin{tabular}{@{}lcc@{}}
\toprule
& Steam & ML-25M \\
\midrule
Users & 281{,}354 & 162{,}541 \\
Items & 11{,}961 & 32{,}720 \\
Item-table size & 11{,}963 & 32{,}722 \\
Training examples & 2{,}706{,}396 & 3{,}226{,}917 \\
Training prefixes per user & All & 20 most recent \\
Embedding source & \texttt{all-MiniLM-L6-v2} (384-d) & \texttt{all-MiniLM-L6-v2} (384-d) \\
Hidden size $D$ & 64 & 64 \\
Split & Leave-one-out & Leave-one-out \\
Test examples & 281{,}354 & 162{,}541 \\
Avg.\ raw seq.\ length & 12.6 & 153.5 \\
Model history limit & 50 & 50 \\
Training batch size & 1{,}024 & 1{,}024 \\
Evaluation batch size & 1{,}024 (DIN and TA-Transformer: 256) & 1{,}024 (DIN and TA-Transformer: 256) \\
\bottomrule
\end{tabular}
\end{table*}

\subsection*{Public Dataset Protocol}

\noindent\textbf{Data and splits.} Both datasets use chronological leave-one-out: the last two items are the validation and test positives, and earlier items form the training sequence. Steam is built from the review dump of \citet{kang2018sasrec} and contains 281{,}354 users and~11{,}961 items across~3{,}550{,}458 filtered interactions. We collapse consecutive repeats before iterative user--item $5$-core filtering. \mbox{ML-25M} has~162{,}541 users and 32{,}720 items after preprocessing. Histories are right-padded or truncated to the $50$ most recent items. Each retained training prefix gives a (history, positive next item) example. We retain all Steam prefixes (2{,}706{,}396 examples) and each user's latest~$20$ \mbox{ML-25M} prefixes (3{,}226{,}917).

\noindent\textbf{Model inputs.} Both datasets use frozen 384-dimensional item-text embeddings from the \texttt{all-\allowbreak MiniLM-\allowbreak L6-v2} checkpoint~\citep{sentenceTransformers2022allMiniLML6v2} and a trainable projection to $D{=}64$. The checkpoint uses the Sentence-BERT framework~\citep{reimers2019sentencebert} with a MiniLM encoder~\citep{wang2020minilm}. Steam items absent from its metadata dump ($<\!1\%$) use placeholder text because all-zero frozen embeddings cannot distinguish or rank them.

\noindent\textbf{Baselines and configurations.} For the public datasets, TA-Transformer has one branch for the single interaction history. It appends the candidate, applies a one-layer, one-head bidirectional Transformer, and scores the candidate-position output. It otherwise shares the common embedding pipeline and uses BPR rather than the industrial model's pointwise binary cross-entropy. The public data contain no unknown item identifiers, so the learned unknown-token parameter is omitted. TA-Transformer has 61{,}569 trainable parameters excluding the frozen item table, and its fixed depth and head count match the industrial comparator. The public \M{} configuration uses four heads ($H{=}4$) of width $d{=}64$ ($D_u{=}256$) and sets $r{=}32$, while the evaluated configurations vary $K\in\{4,8\}$ and cross depth in $\{1,2\}$.

\noindent\textbf{Model size and configurations.} Tab.~\ref{tab:params} lists trainable parameters and the configurations trained on each dataset. The prespecified \M{} configuration has $132{,}885$ trainable parameters, $2.2\times$ TA-Transformer's $61{,}569$ and $2.5\times$ DIN's $52{,}802$. The two SASRec configurations differ only in target normalisation. The three TA-Transformer configurations are the baseline and the two larger variants of Supp.~Sec.~\ref{app:capacity-ta}, with feed-forward widths $2D$ and $4D$. The thirteen \M{} configurations are the four grid cells of Tab.~\ref{tab:main-full} and nine component variants (Tabs.~\ref{tab:ablations} and~\ref{tab:experts}). The headline row is the fixed configuration of Sec.~\ref{sec:public-evaluation}, whereas selecting by validation MRR would choose the 8-expert, 2-layer variant. All reported runs share the optimisation settings below, and our DIEN implementation adds the auxiliary next-behaviour loss~\citep{zhou2019dien} with weight $1$.

\noindent\textbf{Training and model selection.} Training uses BPR with one negative per example, Adam~\citep{kingma2014adam} at learning rate $10^{-3}$, dropout~\citep{srivastava2014dropout} of~$0.1$ after history LN, and batch size 1{,}024. Early stopping monitors validation MRR with patience $10$ for at most $200$ epochs. Evaluation batches contain 1{,}024 examples, or 256 for DIN and TA-Transformer. DIN materialises a $[B,S,C,4D]$ activation tensor, whereas TA-Transformer forms a target-augmented sequence for every candidate in the batch.

\noindent\textbf{Evaluation.} Each held-out test item is scored against $99$ unique negatives sampled uniformly without replacement from outside that user's full sequence and shared across models. The resulting test sets contain 281{,}354 examples for Steam and 162{,}541 for \mbox{ML-25M}. For one training run of each model in Tab.~\ref{tab:main} and of the remaining \M{} configurations in Tab.~\ref{tab:main-full}, counting every tie against the ground-truth item leaves every rank unchanged, so no reported value for those checkpoints depends on tie-breaking. For Tab.~\ref{tab:robust}, we count ties against the ground-truth item. Only $0.5\%$ of Steam test items occur earlier in the visible history, so its task is next-item prediction rather than re-ranking an already-seen item.

\subsection*{Industrial Dataset Protocol}

\noindent\textbf{Data and splits.} Both industrial datasets comprise logged requests with later held-out test periods. Industrial-A draws validation data from its training window, whereas Industrial-B uses a later window for validation and testing. Industrial-A has 591{,}286 test rows from 5{,}836 requests across two disjoint anonymised ranking contexts: Segment-1 (232{,}908) and Segment-2 (358{,}378). Industrial-B has 1{,}088{,}853 test rows from 5{,}625 requests, split into 562{,}927 rows with a user identifier and 525{,}926 with one missing or unknown. These segments reflect identifier availability, not history length.

\noindent\textbf{Model inputs.} For each dataset, all models receive the same request-level features and item representations. Sequential models also receive behavioural histories, partitioned by event type and truncated to recent events.

\noindent\textbf{Baselines and configurations.} Each dataset uses its own tabular baseline. TA-Transformer has one event-history branch per event type. Each branch appends the target, applies a one-layer bidirectional Transformer encoder, and uses the target-position output. It has no cross network or direct target input. DIN also has one branch per event type. Its MLP activation ($\mathrm{MLP}([\mathbf{h}, \mathbf{t}, \mathbf{h} \odot \mathbf{t}, \mathbf{h} - \mathbf{t}])$) computes per-item weights. The separately pooled event-history branches enter DCN-v2 target crosses, and a projected target enters the scorer. \M{} follows Sec.~\ref{sec:architecture}. Both datasets use event-history pooling, one similarity cross layer, a target self-cross, $D{=}256$, $H{=}8$, and $r{=}32$. Industrial-A uses $d{=}64$ ($D_u{=}512$), $K{=}8$, and a per-head gated cross between the event-history branches. Industrial-B uses $d{=}32$ ($D_u{=}256$), $K{=}4$, and a DCN-v2 cross between those branches without direct target input.

\noindent\textbf{Training and evaluation.} Both datasets use binary cross-entropy and validation NDCG@20 for checkpoint selection. Fixed-ratio training negatives are non-purchased candidates from the same request. Evaluation ranks held-out logged candidates and treats purchases as positive outcomes.

\end{document}